\documentclass[aoas,nameyear,seceqn,dvips]{arximspdf}
\usepackage{graphicx}

\doi{10.1214/10-AOAS395}
\volume{5}
\issue{2A}
\pubyear{2011}
\firstpage{1020}
\lastpage{1056}

\makeatletter

\def\eqref#1{(\ref{#1})}

\makeatother

\begin{document}
\begin{frontmatter}

\title{A sticky HDP-HMM with application to speaker
diarization\thanksref{T2}}
\runtitle{The sticky HDP-HMM}

\thankstext{T2}{Supported in part by MURIs funded through AFOSR Grant
FA9550-06-1-0324 and ARO Grant W911NF-06-1-0076, by AFOSR under Grant
FA9559-08-1-0180 and
by DARPA IPTO Contract FA8750-05-2-0249.}

\begin{aug}
\author[A]{\fnms{Emily B.} \snm{Fox}\corref{}\ead[label=e1]{fox@stat.duke.edu}},
\author[B]{\fnms{Erik B.} \snm{Sudderth}\ead[label=e2]{sudderth@cs.brown.edu}},
\author[C]{\fnms{Michael I.} \snm{Jordan}\ead[label=e3]{jordan@stat.berkeley.edu}}
\and\break
\author[D]{\fnms{Alan~S.}~\snm{Willsky}\ead[label=e4]{willsky@mit.edu}}

\runauthor{Fox, Sudderth, Jordan and Willsky}

\affiliation{Duke University, Brown University,
University of California, Berkeley and Massachusetts Institute of Technology}

\address[A]{E. B. Fox\\
Department of Statistical Science\\
Duke University\\
Box 90251\\
Durham, North Carolina 27701\\USA\\
\printead{e1}}

\address[B]{E. B. Sudderth\\
Department of Computer Science\hspace*{6pt}\\
Brown University\\
115 Waterman Street, Box 1910\\
Providence, Rhode Island 02912\\USA\\
\printead{e2}}

\address[C]{M. I. Jordan\\
Department of Statistics\\
\quad and Department of EECS \\
University of California, Berkeley\\
427 Evans Hall\\
Berkeley, California 94720\\USA\\
\printead{e3}}

\address[D]{A. S. Willsky\\
Department of EECS\\
Massachusetts Institute\\\quad  of Technology\\
77 Massachusetts Ave, Rm 32-D582\\
Cambridge, Massachusetts 02139\\USA\\
\printead{e4}}
\end{aug}

% HISTORY:
\received{\smonth{4} \syear{2010}}
\revised{\smonth{8} \syear{2010}}

\begin{abstract}
We consider the problem of \emph{speaker diarization}, the problem
of segmenting an audio recording of a meeting into temporal segments
corresponding to individual speakers. The problem is rendered
particularly difficult by the fact that we are not allowed to assume
knowledge of the number of people participating in the meeting. To
address this problem, we take a Bayesian nonparametric approach to
speaker diarization that builds on the hierarchical Dirichlet
process hidden Markov model (HDP-HMM) of Teh et~al. [\textit{J. Amer. Statist. Assoc.} \textbf{101}
(2006) 1566--1581]. Although
the basic HDP-HMM tends to over-segment the audio data---creating
redundant states and rapidly switching among them---we describe an
augmented HDP-HMM that provides effective control over the switching
rate. We also show that this augmentation makes it possible to
treat emission distributions nonparametrically. To scale the
resulting architecture to realistic diarization problems, we develop
a sampling algorithm that employs a~truncated approximation of the
Dirichlet process to jointly resample the full state sequence,
greatly improving mixing rates. Working with a benchmark NIST data
set, we show that our Bayesian nonparametric architecture yields
state-of-the-art speaker diarization results.\looseness=-1
\end{abstract}

\end{frontmatter}

%s1 ###
\section{Introduction}
\label{sec:Introduction}
A recurring problem in many areas of information technology is that
of segmenting a waveform into a set of time intervals that have a
useful interpretation in some underlying domain. In this article we
focus on a particular instance of this problem, namely, the problem
of \emph{speaker diarization}. In speaker diarization, an audio
recording is made of a meeting involving multiple human participants
and the problem is to segment the recording into time intervals
associated with individual speakers [\citet{Wooters07}]. This
segmentation is to be carried out without a priori knowledge of the
number of speakers involved in the meeting; moreover, we do not
assume that we have a priori knowledge of the speech patterns of
particular individuals.

Our approach to the speaker diarization problem is built on the
framework of hidden Markov models (HMMs), which have been a major
success story not only in speech technology but also in many other
fields involving complex sequential data, including genomics,
structural biology, machine translation, cryptanalysis and finance.
An alternative to HMMs in the speaker diarization setting would be
to treat the problem as a changepoint detection problem, but a key
aspect of speaker diarization is that speech data from a~single
individual generally recurs in multiple disjoint intervals. This
suggests a Markovian framework in which the model transitions among
states that are associated with the different speakers.

An apparent disadvantage of the HMM framework, however, is that
classical treatments of the HMM generally require the number of
states to be fixed a priori. While standard parametric model
selection methods can be adapted to the HMM, there is little
understanding of the strengths and weaknesses of such methods in
this setting, and practical applications of HMMs generally fix the
number of states using ad hoc approaches. It is not clear how to
adapt HMMs to the diarization problem where the number of speakers
is unknown.

Building on the work of \citet{Beal02}, \citet{Teh06} presented
a Bayesian nonparametric version of the HMM in which a stochastic
process---the \textit{hierarchical Dirichlet process} (HDP)---defi\-nes
a prior \mbox{distribution} on transition matrices over countably infinite
state spaces. The resulting \textit{HDP-HMM} is amenable to full
Bayesian posterior inference over the number of states in the model.
Moreover, this posterior distribution can be integrated over when
making predictions, effectively averaging over models of varying
complexity. The HDP-HMM has shown promise in a~variety of applied
problems, including visual scene recognition~[\citet{Kivinen07}],
music synthesis~[\citet{Hoffman08}], and the modeling of genetic
recombination~[\citet{Xing07}] and gene expression~[\citet
{Beal06}].

%f1 ###
\begin{figure*}[t]

\includegraphics{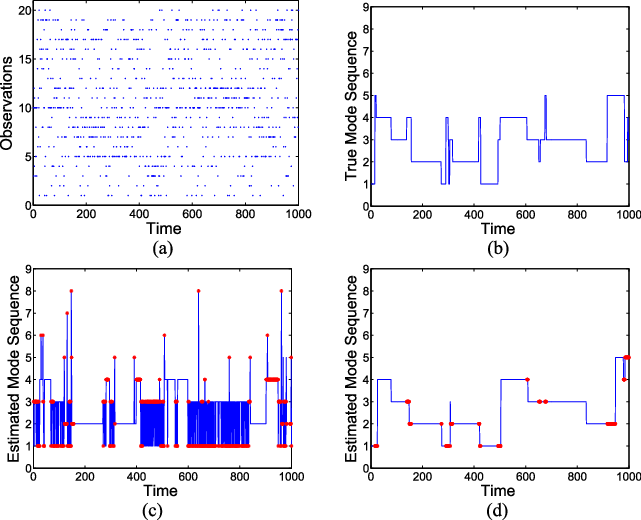}

\caption{\textup{(a)} Multinomial observation sequence; \textup{(b)} true
state sequence; \textup{(c)} and \textup{(d)} estimated state sequence after 30,000 Gibbs
iterations for the original and sticky HDP-HMM, respectively, with
errors indicated in red. Without an extra self-transition bias, the
HDP-HMM rapidly transitions among redundant states.}
\label{fig:rapid_dynamics}
\end{figure*}

While the HDP-HMM seems like a natural fit to the speaker
diarization problem given its structural flexibility, as we show in
Section~\ref{sec:SpeakerDiarization}, the HDP-HMM does not yield
state-of-the-art performance in the speaker diarization setting.
The problem is that the HDP-HMM inadequately models the temporal
persistence of states. This problem arises in classical finite HMMs
as well, where semi-Markovian models are often proposed as
solutions. However, the problem is exacerbated in the nonparametric
setting, in which the Bayesian bias toward simpler models is
insufficient to prevent the HDP-HMM from giving high posterior
probability to models with unrealistically rapid switching. This is
demonstrated in Figure~\ref{fig:rapid_dynamics}, where we see that the
HDP-HMM sampling algorithm creates redundant states and rapidly
switches among them. (The figure also displays results from the
augmented HDP-HMM---the ``sticky HDP-HMM'' that we describe in this
paper.)
The tendency to create redundant states is not necessarily a problem
in settings in which model averaging is the goal. For speaker
diarization, however, it is critical to infer the number of speakers
as well as the transitions among speakers.\looseness=-1

Thus, one of our major goals in this paper is to provide a general
solution to the problem of state persistence in HDP-HMMs. Our
approach is easily stated---we simply augment the HDP-HMM to include
a parameter for self-transition bias, and place a separate prior on
this parameter. The challenge is to execute this idea coherently in
a Bayesian nonparametric framework. Earlier papers have also
proposed self-transition parameters for HMMs with infinite state
spaces~[\citet{Beal02}; \citet{Xing07}], but did not formulate general
solutions that integrate fully with Bayesian nonparametric
inference.

Another goal of the current paper is to develop a more fully
nonparametric version of the HDP-HMM in which not only the
transition distribution but also the emission distribution (the
conditional distribution of observations given states) is treated
nonparametrically. This is again motivated by the speaker
diarization problem---in classical applications of HMMs to speech
recognition problems, it is often the case that emission
distributions are found to be multimodal, and high-performance HMMs
generally use finite Gaussian mixtures as emission
distributions [\citet{GalesYoung}]. In the nonparametric setting
it is
natural to replace these finite mixtures with Dirichlet process
mixtures. Unfortunately, this idea is not viable in practice,
because of the tendency of the HDP-HMM to rapidly switch between
redundant states. As we show, however, by incorporating an
additional self-transition bias, it is possible to make use of
Dirichlet process mixtures for the emission distributions.

An important reason for the popularity of the classical HMM is its
computational tractability. In particular, marginal probabilities
and samples can be obtained from the HMM via an efficient dynamic
programming algorithm known as the forward--backward
algorithm~[\citet{Rabiner89}]. We show that this algorithm also
plays an important role in computationally efficient inference for
our generalized HDP-HMM. Using a truncated approximation to the full
Bayesian nonparametric model, we develop a blocked Gibbs
sampler which leverages forward--backward recursions to jointly
resample the state and emission assignments for all observations.

The paper is organized as follows. In Section~\ref
{sec:SpeakerDiarizationTask} we begin
by summarizing related prior work on the speaker diarization task and
analyzing the key characteristics of the data set we examine in
Section~\ref{sec:SpeakerDiarization}. In Section~\ref{sec:DP} we
provide some basic background on Dirichlet processes.
Then, in Section~\ref{sec:HDP} we overview the hierarchical
Dirichlet process, and in Section~\ref{sec:stickyHDPHMM} discuss how
it applies to HMMs and can be extended to account for state
persistence. An efficient Gibbs sampler is also described in this
section. In Section~\ref{sec:temperedHDPHMMDP} we treat the case of
nonparametric emission distributions. We discuss our application to
speaker diarization in Section~\ref{sec:SpeakerDiarization}. A list of
notational conventions can be found in the Supplementary
Material~[\citet{SuppA}].

%s2 ###
\section{The speaker diarization task}
\label{sec:SpeakerDiarizationTask}
There is a vast literature on the speaker diarization task, and in this
section we simply aim to provide an overview of the most common
techniques. We refer the interested reader to \citet{Tranter06}
for a more thorough exposition on the subject.

Classical speaker diarization techniques typically employ a two-stage
procedure that first segments the audio (or features thereof) using one
of a variety of changepoint algorithms. The inferred segments are then
regrouped into a set of speaker labels via a clustering algorithm. For
example, \citet{Reynolds04} propose a changepoint detection method
based on the Bayesian Information Criterion (BIC). Specifically, a
penalized likelihood ratio test is used to compare whether the data
within a fixed window are better modeled via a single Gaussian or two
Gaussians. The window gradually grows at each test until a changepoint
is inferred, at which point the window is reinitialized at the inferred
changepoint. An alternative changepoint detection technique, first
proposed in \citet{Siegler97}, uses fixed length windows and
computes the symmetric Kullback--Leibler (KL) divergence between a pair
of Gaussians each fit by the data in their respective windows. A
post-processing step then sets the changepoints equal to the peaks of
the computed KL that exceed a predetermined threshold. In order to
group\vadjust{\goodbreak} the inferred segments into a set of speaker labels, a common
approach is to use hierarchical agglomerative clustering with a BIC
stopping criterion, as proposed in \citet{Chen98}.

The simple two-stage approach outlined above suffers from the fact that
errors made in the segmentation stage can degrade the performance of
the subsequent clustering stage. A number of algorithms instead iterate
between multiple stages of resegmentation (typically via Viterbi
decoding) and clustering; for example, see \citet
{Barras04}; \citet{Wooters04}. Iterative segmentation and clustering
algorithms employing a Gaussian mixture model for each cluster (i.e.,
speaker), such as those proposed by \citet{Gauvain98}; \citet{Barras04}, have
been shown to improve diarization performance. Overall, however,
agglomerative clustering is extremely sensitive to the specified
threshold for cluster merging, with different settings leading to
either over- or under-clustering of the segments into speakers. The
thresholds are typically set based on testing on an extensive training database.

A number of more recent approaches have considered the problem of joint
segmentation and clustering by employing HMMs to capture the repeated
returns of speakers. To handle the fact that the state space is
unknown, \citet{Meignier00} introduces the use of an evolutive-HMM
which is further developed in \citet{Meignier01}. The HMM is
initialized to have one state and at each iteration a segment of speech
is assumed to arise from an undetected speaker who is added to the
model. The revised HMM is then used to resegment the audio, and this
iterative procedure continues until the speaker labels have converged.
An alternative HMM formulation is presented in \citet
{Wooters07}. The data are initially split into $K$ states, with $K$
assumed to be larger than the number of true speakers, and the HMM
states are iteratively merged according to a metric based on changes in
BIC. At each iteration, Viterbi decoding is performed to resegment the
features of the audio, and the inferred segments are used to fit a new
HMM via expectation maximization (EM). Then, the BIC criterion is
applied to decide whether to merge HMM states. The algorithm also
includes HMM substates to impose minimum speaker durations.

Our approach also seeks to jointly segment and cluster the audio into~%
spea\-ker-homogenous regions, as targeted by the HMM approaches
of \citet{Meignier01}; \citet{Wooters07}, but within a Bayesian
nonparametric framework that avoids relying on the heuristics employed
by these previously proposed algorithms and allows for coherent
Bayesian inference.

The data set we consider in the experiments of Section~\ref
{sec:SpeakerDiarization} is a standard benchmark data set
distributed by NIST as part of the Rich Transcription 2004--2007
meeting recognition evaluations [\citet{NIST}]. The data set
consists of 21 recorded meetings, each of which may have different sets
of speakers both in number and identity. We use the first 19
Mel Frequency Cepstral\vadjust{\goodbreak} Coefficients (MFCCs),\footnote{Mel-frequency
cepstral coefficients (MFCCs) comprise a representation of the
short-term power spectrum of a sound on the mel scale (a nonlinear
scale of frequency based on the human auditory system response).
Specifically, the computation of an MFCC typically involves (i) taking
the Fourier transform of a windowed excerpt of a signal, (ii) mapping
the log powers of the obtained spectrum onto the mel scale and (iii)
performing a discrete cosine transform of the mel log powers. The MFCCs
are the amplitudes of the resulting spectrum.} computed over a 30~ms
window every 10~ms, as a~feature vector. After these features are
computed, a speech/nonspeech
detector is run to identify and remove observations corresponding to
nonspeech. (Nonspeech refers to
time intervals in which nobody is speaking.) The preprocessing
step of removing nonspeech observations is important in ensuring
that the fitted acoustic models are not corrupted by nonspeech
information.\looseness=-1

\begin{figure*}[t]

\includegraphics{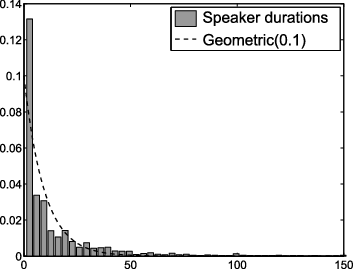}
\vspace*{-5pt}
\caption{Normalized histogram of speaker durations of the preprocessed
audio features from the 21 meetings in the NIST database. A $\operatorname
{Geom}(0.1)$ density is also shown for comparison.}
\label{fig:SpeakerDuration}
\vspace*{-5pt}
\end{figure*}

When working with this data set, we discovered that the high frequency
content of these
features contained little discriminative information. Since minimum
speaker durations are rarely less than 500~ms, we chose to define the
observations as
averages over 250~ms, nonoverlapping blocks. This preprocessing stage
also aids in achieving speaker dynamics at the correct granularity (as
opposed to finer temporal scale features leading to inferring
within-speaker dynamics in addition to global speaker changes). In
Figure~\ref{fig:SpeakerDuration} we plot a histogram of the speaker
durations of our preprocessed features based on the ground truth labels
provided for each of the 21 meetings. From this plot, we see that a
geometric duration distribution fits this data reasonably well. This
motivates our approach of simply increasing the prior probability of
self-transitions within a Markov framework rather than moving to the
more complicated semi-Markov formulation of speaker transitions.

%f2 ###

%f3 ###
\begin{figure*}[b]
\vspace*{-5pt}
\includegraphics{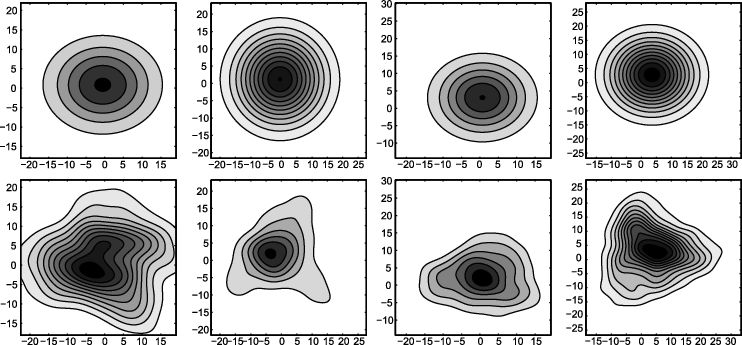}%
\vspace*{-5pt}
\caption{Contour plots of the best fit Gaussian (top) and kernel
density estimate (bottom) for the top two principal components of the
audio features associated with each of the four speakers present in the
AMI\_20041210-1052 meeting. Without capturing the non-Gaussianity of
the speaker-specific emissions, the speakers are challenging to identify.}
\label{fig:SpeakerDiarizationPCA}
\end{figure*}

Another key feature of the speaker diarization data is the fact that
the speaker specific emissions are not well approximated by a single
Gaussian; see Figure~\ref{fig:SpeakerDiarizationPCA}. This observation
has led many researchers to consider a mixture-of-Gaussians speaker
model, as previously described. As demonstrated in Section~\ref
{sec:SpeakerDiarization}, we show that achieving state-of-the-art
performance within our framework also relies on allowing for
non-Gaussian emissions.\vadjust{\goodbreak}

%s3 ###
\section{Dirichlet processes}
\label{sec:DP}

A Dirichlet process (DP) is a distribution on probability measures
on a measurable space $\Theta$. This stochastic process is uniquely
defined by a base measure $H$ on $\Theta$ and a concentration
parameter $\gamma$; we denote it by $\operatorname{DP}({\gamma
},{H})$. Consider a
random probability measure $G_0 \sim\operatorname{DP}({\gamma
},{H})$. The DP is
formally defined by the property that, for any finite
partition %\footnote{A partition of a set $A$ is a set of disjoint,
%non-empty subsets of $A$ such that every element of $A$ is contained
%in exactly one of these subsets. More formally, $\{A_k\}_{k=1}^K$ is
%a partition of $A$ if $\cup_k A_k = A$ and for each $j\neq k$, $A_k
$\{A_1,\dots,A_K\}$ of $\Theta$,\vspace*{-1pt}
%
%e3.1 ###
\begin{eqnarray}\label{eqn:finite_partition}
(G_0(A_1),\dots,G_0(A_K))|\gamma,H \sim\operatorname{Dir}(\gamma
H(A_1),\dots,\gamma H(A_K)).\vspace*{-1pt}
\end{eqnarray}
That is, the measure of a random probability distribution $G_0\sim
\operatorname{DP}({\gamma},{H})$ on every finite partition of $\Theta
$ follows a
finite-dimensional Dirichlet \textit{distribution} [\citet{Ferguson73}].
%This definition of the DP is due to %Ferguson \citet{Ferguson73}, who
%invoked Kolmogorov's consistency conditions to establish the existence
%of the DP as a stochastic process with Dirichlet marginals.
A more constructive definition of the DP was
given by %Sethuraman
\citet{Sethuraman94}. Consider a probability mass function (p.m.f.)
$\{\beta_k\}_{k=1}^\infty$ on a countably infinite set, where the
discrete probabilities are defined as follows:\vspace*{-3pt}
\begin{eqnarray}
v_k|\gamma&\sim&\operatorname{Beta}(1,\gamma), \qquad
k=1,2,\dots,\nonumber
\\[-11pt]\\[-11pt]
\beta_k &=& v_k\prod_{\ell=1}^{k-1}(1-v_\ell), \qquad k=1,2,\dots.\nonumber\vspace*{-3pt}
\end{eqnarray}
In effect, we have divided a unit-length stick into lengths given by
the weights~$\beta_k$: the $k${th} weight is a random proportion
$v_k$ of the remaining stick after the previous $(k-1)$ weights have
been defined. This \textit{stick-breaking construction} is generally
denoted by $\beta\sim\operatorname{GEM}(\gamma)$. With probability one,
a~random draw $G_0 \sim \operatorname{DP}(\gamma,H)$ can be expressed as\vspace*{-3pt}
%
%e3.2 ###
\begin{eqnarray}\label{eqn:stickDP}
G_0 = \sum_{k=1}^\infty\beta_k\delta_{\theta _k},
\qquad \theta _k | H \sim H,  k = 1,2,\dots,\vspace*{-3pt}
\end{eqnarray}
where $\delta_{\theta}$ denotes a unit-mass measure concentrated at
$\theta$ and where $\{\theta_k\}$ are drawn independently from $H$.
From this definition, we see that the DP actually defines
a distribution over discrete probability measures. The
stick-breaking construction also gives us insight into how the
concentration parameter $\gamma$ controls the relative magnitude of
the mixture weights $\beta_k$, and thus determines the model
complexity in terms of the expected number of components with
significant probability mass.

The DP has a number of properties which make inference based on this
nonparametric prior computationally tractable. Consider a set of
observations $\{\theta'_i\}$ with $\theta'_i\sim G_0$. Because
probability measures drawn from a DP are discrete, there is a
strictly positive probability of multiple observations~$\theta'_i$
taking identical values within the set $\{\theta _k\}$, with
$\theta _k$ defined as in equation~\eqref{eqn:stickDP}. For each
value $\theta'_i$, let $z_i$ be an indicator random variable that
picks out the unique value $k$ such that $\theta'_i =
\theta _{z_i}$. %Blackwell and MacQueen
\citet{Blackwell73} introduced a P\'{o}lya urn representation of
the $\theta'_i$:\vspace*{-3pt}
\begin{eqnarray}\label{eqn:PolyaUrn}
\theta'_{i} |\theta'_1, \dots, \theta'_{i-1} &\sim&
\frac{\gamma}{\gamma+ i-1}H + \sum_{j=1}^{i-1} \frac{1}{\gamma+
i-1}\delta_{\theta'_j}\nonumber
\\[-11pt]\\[-11pt]
&=& \frac{\gamma}{\gamma+ i-1}H + \sum_{k=1}^{K}
\frac{n_k}{\gamma+ i-1}\delta_{\theta _k},\nonumber\vspace*{-3pt}
\end{eqnarray}
implying the following predictive distribution for the indicator
random variables:\vspace*{-3pt}
%
%e3.3 ###
\begin{eqnarray}\label{eqn:predictive}
\qquad\ \ p(z_{N+1} = z| z_1,\dots,z_N,\gamma)
= \frac{\gamma}{N+\gamma}
\delta(z,K+1)+ \frac{1}{N+\gamma}\sum_{k=1}^K n_k\delta(z,k).\vspace*{-3pt}
\end{eqnarray}
Here, $n_k=\sum_{i=1}^N \delta(z_i,k)$ is the number of indicator
random variables taking the value $k$, and $K+1$ is a previously
unseen value. We use the\vadjust{\eject} notation~$\delta(z,k)$ to indicate the
discrete Kronecker delta. This representation can be used to sample
observations from a DP without explicitly constructing the countably
infinite random probability measure $G_0 \sim\operatorname
{DP}({\gamma},{H})$.

%f4 ###
\begin{figure*}[t]

\includegraphics{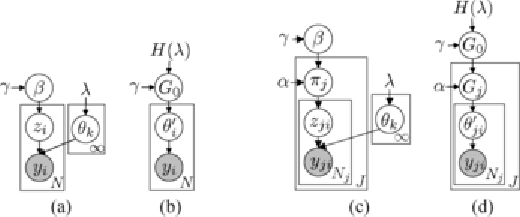}%
\vspace*{-5pt}
\caption{Dirichlet
process \textup{(left)} and hierarchical Dirichlet process \textup{(right)}
mixture models represented in two different ways as graphical
models. \textup{(a)} Indicator variable representation in which $\beta|\gamma
\sim\operatorname{GEM}(\gamma)$, $\theta _k|H,\lambda\sim
H(\lambda)$, $z_i|\beta\sim\beta$ and
$y_i|\{\theta _k\}_{k=1}^\infty,z_i \sim
F(\theta _{z_i})$. \textup{(b)} Alternative representation with
$G_0|\gamma,H \sim\operatorname{DP}({\gamma},{H})$, $\theta'_i|G_0
\sim G_0$, and
$y_i|\theta'_i \sim F(\theta'_i)$. \textup{(c)} Indicator variable
representation in which $\beta|\gamma\sim\operatorname{GEM}(\gamma)$,
$\pi_k|\alpha,\beta\sim\operatorname{DP}({\alpha},{\beta})$,
$\theta _k|H,\lambda\sim H(\lambda)$, $z_{ji}|\pi_j \sim
\pi_j$, and $y_{ji}|\{\theta _k\}_{k=1}^\infty,z_{ji} \sim
F(\theta _{z_{ji}})$. \textup{(d)} Alternative representation with
$G_0|\gamma,H \sim\operatorname{DP}({\gamma},{H})$, $G_j|G_0 \sim
\operatorname{DP}({\alpha},{G_0})$,
$\theta'_{ji}|G_j \sim G_j$ and $y_{ji}|\theta'_{ji} \sim F(
\theta'_{ji})$. The ``plate'' notation is used to compactly
represent replication [\protect\citet{Teh06}].} \label{fig:DPMM}
\vspace*{-6pt}
\end{figure*}

The distribution on partitions induced by the sequence of
conditional distributions in equation~\eqref{eqn:predictive} is commonly
referred to as the \textit{Chinese restaurant process}. The analogy,
which is useful in developing various generalizations of the
Dirichlet process we consider in this paper, is as follows. Take~%
$i$ to be a customer entering a restaurant with infinitely
many tables, each serving a unique dish $\theta _k$. Each
arriving customer chooses a table, indicated by $z_i$, in proportion
to how many customers are currently sitting at that table. With some
positive probability proportional to $\gamma$, the customer starts a
new, previously unoccupied table $K+1$. The Chinese restaurant process
captures the fact that the DP has a clustering property such that
multiple draws from the random measure take the same value.
%
%From Eq.~\eqref{eqn:predictive} we see that when $z_i \sim\beta$
%and $\beta\sim\operatorname{GEM}(\gamma)$, we can integrate out $\beta$ to
%determine a closed-form predictive distribution for $z_i$. We can
%also find the distribution of the number of unique values of $z_i$
%resulting from $N$ draws from the measure $\beta$. Letting $K$ be
%the number of unique values of $\{z_1,\dots,z_N\}$, this
%distribution is given by \citep{Antoniak74}:
%%
%p(K\mid N,\gamma) = \frac{\Gamma(\gamma)}{\Gamma(\gamma+
%N)}s(N,K)\gamma^K, \label{eqn:unique_stick}
%%
%where $s(n,m)$ are unsigned Stirling numbers of the first
%kind.%~\citep{Abromowitz:72}.

The DP is commonly used as a prior on the parameters of a mixture
model with a random number of components. Such a model is called a
\textit{Dirichlet process mixture model} and is depicted as a
graphical model in Figure~\ref{fig:DPMM}(a) and (b). To generate
observations, we choose $\theta'_i \sim G_0$ and $y_i \sim
F(\theta'_i)$ for an indexed family of distributions $F(\cdot)$.
This sampling process is also often described in terms of the
indicator random variables $z_i$; in particular, we have $z_i \sim
\beta$ and $y_i \sim F(\theta _{z_i})$. The parameter with
which an observation is associated implicitly partitions or clusters
the data. In addition, the Chinese restaurant process representation
indicates that the DP provides a prior that makes it more likely to
associate an observation with a parameter to which other
observations have already been associated. This reinforcement
property is essential for inferring finite, compact mixture models.
It can be shown under mild conditions that if the data were
generated by a finite mixture, then the DP posterior is guaranteed
to converge (in distribution) to that finite set of mixture
parameters [\citet{Ishwaran02}].

%Finally, we can also obtain the DP mixture model as the limit of a
%sequence of finite mixture models. Let us assume that there are $L$
%components in a finite mixture model and we place a
%finite-dimensional Dirichlet prior on these mixture weights:
%%
%%
%%%
%%\begin{eqnarray}
%%\begin{array}{ll}
%%\beta\sim\operatorname{Dir}(\gamma/L,\dots,\gamma/L) & z_i \sim\beta\\
%%\theta_k \sim H(\lambda) & y_i \sim F(\theta_{z_i}).
%%\end{array}
%%\end{eqnarray}
%%%
%Let $G_0^L = \sum_{k=1}^L \beta_k \delta_{\uniqueTheta_k}$. Then, it
%can be shown~\citep{Ishwaran02-2,Ishwaran00} that for every
%measurable function $f$ integrable with respect to the measure $H$,
%this finite distribution $G_0^L$ converges weakly to a countably
%infinite distribution $G_0$ distributed according to a Dirichlet
%process. That is, as $L \rightarrow\infty$,
%%
%%
%%as $L \rightarrow\infty$ for $G_0 \sim DP(\gamma,H)$.
%

%s4 ###
\section{Hierarchical Dirichlet processes}
\label{sec:HDP}
In the following section we describe how ideas based on the Dirichlet process
have been used to develop a~Bayesian nonparametric approach to hidden Markov
modeling in which the number of states is unknown a priori. To develop this
nonparametric version of the HMM, the Dirichlet process does not suffice;
rather, it is necessary to develop a hierarchical Bayesian model involving
a tied collection of Dirichlet processes. This has been done by
\citet{Teh06}
whose \textit{hierarchical Dirichlet process (HDP)} we describe in this section.
The HDP is applicable to general problems involving related groups of data,
each of which can be modeled using a DP, and we begin by describing the HDP
at this level of generality, subsequently specializing to the HMM.

To describe the HDP, suppose there are $J$ groups of data and
let $\{y_{j1},\dots,\break y_{j{N_j}}\}$ denote the set of observations
in group $j$. Assume that there are a~collection of DP mixture
models underlying the observations in these groups:
%
%e4.1 ###
\begin{eqnarray}
\label{eqn:HDPMM}
G_j &=& \sum_{t=1}^\infty\tilde{\pi}_{jt} \delta_{\theta^{*}_{jt}},
\qquad \tilde{\pi}_j|\alpha\sim\operatorname{GEM}(\alpha),
 j=1,\ldots,J,\nonumber
\\[-2pt]
&&\qquad\hspace*{51pt} \theta^{*}_{jt}| G_0, \sim G_0, t=1,2,\ldots,
\\[-2pt]
\qquad \theta'_{ji}|  G_j &\sim& G_j, \qquad y_{ji}| \theta'_{ji} \sim
F(\theta'_{ji}),\qquad j=1,\dots,J,
i=1,\dots,N_j.\nonumber
\end{eqnarray}
We wish to tie the DP mixtures across the different groups such that
atoms that underly the data in group $j$ can be used in group $j^\prime$.
The problem is that if $G_0$ is absolutely continuous with respect
to the Lebesgue measure (as it generally is for continuous parameters),
then the atoms in $G_j$ will be distinct from those in $G_{j^\prime}$
with probability one. The solution to this problem is to let $G_0$
itself be a draw from a DP:
\begin{eqnarray}\label{eqn:HDPglobaldist}
G_0 = \sum_{k=1}^\infty\beta_k \delta_{\theta _k},\qquad\hspace*{-4pt}  &&\beta|
\gamma\sim\operatorname{GEM}(\gamma),\nonumber
\\[-11pt]\\[-11pt]
&& \theta _k|  H,\lambda\sim H(\lambda),  k=1,2,\dots.\nonumber
\end{eqnarray}
In this hierarchical model, $G_0$ is atomic and random. Letting
$G_0$ be a base measure for the draw $G_j \sim\operatorname
{DP}({\alpha},{G_0})$
implies that only these atoms can appear in $G_j$. Thus, atoms can
be shared among the collection of random measures $\{G_j\}$.
The HDP model is depicted graphically in two different ways in
Figure~\ref{fig:DPMM}(c) and (d).

\citet{Teh06} have also described the marginal probabilities
obtained from integrating over the random measures $G_0$ and $\{G_j\}$.
They show that these marginals can be described in terms of a
\textit{Chinese restaurant franchise} (CRF) that is an analog of the
Chinese restaurant process. The CRF is comprised of $J$ restaurants,
each corresponding to an HDP group, and an infinite buffet line of
dishes common to all restaurants. The process of seating customers
at tables, however, is restaurant specific. Each customer is
preassigned to a given
restaurant determined by that customer's group $j$. Upon entering the $j${th}
restaurant in the CRF, customer $y_{ji}$ sits at currently occupied
tables $t_{ji}$ with probability proportional to the number of
currently seated customers, or starts a new table $T_j + 1$ with
probability proportional to $\alpha$. The first customer to sit at a
table goes to the buffet line and
picks a~dish~$k_{jt}$ for their table, choosing the dish with
probability proportional to the number of times that dish has been
picked previously, or ordering a~new dish $\theta _{K+1}$ with
probability proportional to $\gamma$. The intuition behind this
predictive distribution is that integrating over the global dish
probabilities~$\beta$ results in customers making decisions based on the observed
popularity of the dishes throughout the entire franchise. See the
Supplementary Material for further details~[\citet{SuppA}].

Recalling equations~\eqref{eqn:HDPMM} and \eqref{eqn:HDPglobaldist},
since each distribution $G_j$ is drawn from a DP with a discrete
base measure $G_0$, multiple $\theta^{*}_{jt}$ may take an
identical value\vspace*{1pt} $\theta _k$ for multiple unique values of $t$. As
we see in the Supplemental Material~[\citet{SuppA}], this
corresponds to multiple tables in the same restaurant being served
the same dish. %, as depicted in Fig.~\ref{fig:CRFtables}.
We can write $G_j$ as a function of the unique dishes:\vspace*{-3pt}
%
%e4.2 ###
\begin{eqnarray}
G_j = \sum_{k=1}^\infty\pi_{jk} \delta_{\theta _k}, \qquad
\pi_j| \alpha,\beta\sim\operatorname{DP}({\alpha},{\beta}),
\theta _k
|  H \sim H,\vspace*{-3pt}
\end{eqnarray}
where $\pi_j$ now defines a restaurant-specific distribution over
dishes served rather than over tables, with\vspace*{-3pt}
%
%e4.3 ###
\begin{eqnarray}
\pi_{jk} = \sum_{t| \theta^{*}_{jt}=\theta _k} \tilde{\pi}_{jt}.\vspace*{-3pt}
\end{eqnarray}
%
%%
% \centering
% \includegraphics[height=1.75in]{JournalFigs/CRFtables}
% \caption{CRF with $J=2$ restaurants. The currently
%occupied tables each choose a dish $\tableTheta_{jt}|G_j \sim G_j$,
%where $G_j|G_0 \sim\DP{\alpha}{G_0}$ is a discrete probability
%measure so that multiple tables may serve the same dish. Since $G_1$
%has overlapping support with $G_2$, parameters (i.e., dishes) are
%shared between restaurants.} \label{fig:CRFtables}
%%

Let $z_{ji}$ be the indicator random variable for the unique dish
selected by observation $y_{ji}$. An
 equivalent representation for the generative model is in terms
of these indicator random variables:
%
%e4.4 ###
\begin{eqnarray}
\quad \pi_j| \alpha,\beta\sim\operatorname{DP}({\alpha},{\beta}), \qquad
z_{ji}| \pi_j \sim\pi_j, \qquad y_{ji}|
\{\theta _k\},z_{ji} \sim F(\theta _{z_{ji}}),
\end{eqnarray}
and is shown in Figure~\ref{fig:DPMM}(c).

%As with the DP, the HDP mixture model has an interpretation as the
%limit of a finite mixture model. Placing a finite Dirichlet prior on
%$\beta$ induces a finite Dirichlet prior on $\pi_j$:
%%
%%
%As $L\rightarrow\infty$, this model converges in distribution to
%the HDP mixture model \citep{Teh06}.

%s5 ###
\section{The sticky HDP-HMM}
\label{sec:stickyHDPHMM}
Recall that the hidden Markov model, or \textit{HMM}, is a class of
doubly stochastic processes based on an underlying, discrete-valued
state sequence, which is modeled as Markovian [\citet{Rabiner89}].
Let $z_t$ denote the state of the Markov chain at time~$t$ and
$\pi_j$ the state-specific transition distribution for state $j$.
Then, the Markovian structure on the state sequence dictates that
$z_t \sim\pi_{z_{t-1}}$. The observations, $y_t$, are conditionally
independent given this state sequence, with $y_t \sim
F(\theta _{z_t})$ for some fixed distribution $F(\cdot)$.

\begin{figure*}[t]

\includegraphics{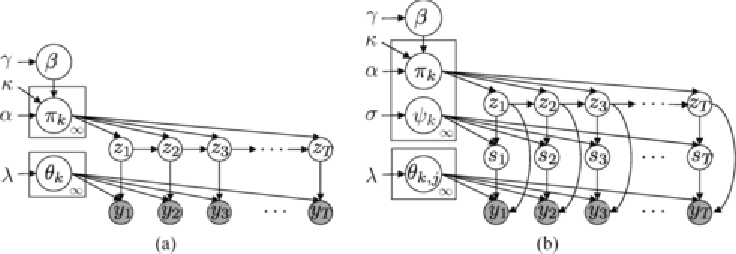}%
\vspace*{-4pt}
\caption{\textup{(a)} Graphical representation of the sticky
HDP-HMM. The state evolves as $z_{t+1}|\{\pi_k\}_{k=1}^{\infty},z_t
\sim\pi_{z_t}$, where $\pi_k|\alpha,\kappa,\beta\sim
\operatorname{DP}(\alpha+\kappa,
(\alpha\beta+\kappa\delta_k)/(\alpha+\kappa))$ and $\beta|\gamma
\sim\operatorname{GEM}(\gamma)$, and observations are generated as
$y_t|\{\theta _k\}_{k=1}^{\infty},z_t \sim
F(\theta _{z_t})$. The original HDP-HMM has $\kappa=0$. \textup{(b)}
Sticky HDP-HMM with DP emissions, where $s_t$ indexes the
state-specific mixture component generating observation $y_t$. The
DP prior dictates that $s_t|\{\psi_k\}_{k=1}^{\infty},z_t \sim
\psi_{z_t}$ for $\psi_k|\sigma\sim\operatorname{GEM}(\sigma)$. The
$j${th} Gaussian component of the $k${th} mixture density is
parameterized by $\theta _{k,j}$ so
$y_t|\{\theta _{k,j}\}_{k,j=1}^{\infty},z_t,s_t \sim
F(\theta _{z_t,s_t})$.}\label{fig:HDPHMM}\vspace*{-5pt}
\end{figure*}

The HDP can be used to develop an HMM with an infinite state
space---the HDP-HMM~[\citet{Teh06}]. In the speaker diarization
task, each state constitutes a different speaker and our goal in moving
to an infinite state space is to remove upper bounds on the total
number of speakers present. Conceptually, we envision a
doubly-infinite transition matrix, with each row corresponding to a
Chinese restaurant. That is, the groups in the HDP formalism here
correspond to states, and each Chinese restaurant defines a~distribution
on next states. The CRF links these next-state
distributions. Thus, in this application of the HDP, the
group-specific distribution, $\pi_j$, is a state-specific transition
distribution and, due to the infinite state space, there are
infinitely many such groups. Since $z_t \sim\pi_{z_{t-1}}$, we see
that $z_{t-1}$ indexes the group to which $y_t$ is assigned (i.e.,
all observations with $z_{t-1}=j$ are assigned to group $j$). Just
as with the HMM, the current state $z_t$ then indexes the parameter
$\theta _{z_t}$ used to generate observation $y_t$ [see
Figure~\ref{fig:HDPHMM}(a)].

By defining $\pi_j \sim\operatorname{DP}({\alpha},{\beta})$, the
HDP prior
encourages states to have similar transition distributions
($E[\pi_{jk}| \beta]=\beta_k$). However, it does not
differentiate self-transitions from moves between different states.
%We explore the
%limitations of this construction, and develop a more general
%framework which allows domain-specific learning of self-transition
%biases.
When modeling data with state persistence, the flexible nature of
the HDP-HMM prior allows for state sequences with unrealistically
fast dynamics to have large posterior probability. For example, with
multinomial emissions, a good explanation of the data is to divide
different observation values into unique states and then rapidly
switch between them (see Figure~\ref{fig:rapid_dynamics}).
%
%For example, with Gaussian emissions parameterized by unknown means
%and variances, as in Fig.~\ref{fig:fast_switching}, a good
%explanation of the data is to divide an observation block into two
%small--variance states with slightly different means, and then
%rapidly switch between them (see Fig.~\ref{fig:fast_switching}).
%
In such cases, many models with redundant states may have large
posterior probability, thus impeding our ability to identify a
compact dynamical model which best explains the observations. The
problem is compounded by the fact that once this alternating pattern
has been instantiated by the sampler, its persistence is then
reinforced by the properties of the Chinese restaurant franchise,
thus slowing mixing rates. Furthermore, this fragmentation of data
into redundant states can reduce predictive performance, as is
discussed in Section~\ref{sec:results1}. In many applications, one
would like to be able to incorporate prior knowledge that slow,
smoothly varying dynamics are more likely.

%f5 ###

To address these issues, we propose to instead model the transition
distributions $\pi_j$ as follows:
\begin{eqnarray}\label{eqn:biasedHDPHMM}
\beta| \gamma&\sim&\operatorname{GEM}(\gamma),\nonumber
\\[-10pt]\\[-10pt]
\pi_j| \alpha,\kappa,\beta&\sim&\operatorname{DP}\biggl(\alpha+ \kappa,
\frac{\alpha\beta+ \kappa\delta_j}{\alpha+ \kappa}\biggr).\nonumber
\end{eqnarray}
Here, $(\alpha\beta+ \kappa\delta_j)$ indicates that an amount
$\kappa> 0 $ is added to the $j${th} component of $\alpha\beta$.
Informally, what we are doing is increasing the expected probability
of self-transition by an amount proportional to $\kappa$:
%
%e5.1 ###
\begin{eqnarray}
E[\pi_{jk} | \beta, \kappa] = \frac{\alpha\beta_k + \kappa
\delta(j,k)}{\alpha+\kappa}.
\end{eqnarray}
More formally, over a finite partition $(Z_1,\dots,Z_K)$ of the positive
integers~$\mathbb{Z}_+$, the prior on the measure $\pi_j$ adds an
amount $\kappa$ only to the arbitrarily small partition containing
$j$, corresponding to a self-transition. That is,
\begin{eqnarray}
&&(\pi_j(Z_1),\dots,\pi_j(Z_K))| \alpha,\beta\nonumber
\\[-8pt]\\[-8pt]
&&\qquad \sim
\operatorname{Dir}\bigl(\alpha\beta(Z_1)+\kappa\delta_j(Z_1),\dots,\alpha
\beta(Z_K)+\kappa\delta_j(Z_K)\bigr).\nonumber
\end{eqnarray}
When $\kappa=0$ the original HDP-HMM of \citet{Teh06} is recovered.
Because positive $\kappa$ values increase the prior probability
$E[\pi_{jj} | \beta]$ of self-transitions, we refer to this
extension as the \textit{sticky} HDP-HMM. See
Figure~\ref{fig:HDPHMM}(a). Note that this formulation assumes that
the stickiness of each HMM state is the same a priori. The parameter
could be made state-dependent through a hierarchical model that ties
together a collection of state-specific sticky parameters. However,
such state-specific stickiness is unnecessary for the speaker
diarization task at hand since each speaker is assumed to have similar
expected durations. Differences between speaker-specific transitions
become more distinguished in the posterior.

The $\kappa$ parameter is reminiscent of the self-transition bias
parameter of the \textit{infinite HMM}, an urn model for hidden Markov
models on infinite state spaces that predated the HDP-HMM [\citet{Beal02}].
The connection between the (sticky) HDP-HMM and the infinite HMM is analogous
to that between the DP and the P\'olya urn; in both cases the latter
is obtained by integrating out the random measures in the former.
In particular, the infinite HMM employs a two-level urn model in
which the top-level urn places a probability on transitions to existing
states in proportion to how many times these transitions have been
seen, with an added bias for a self-transition even if this has
not previously occurred. With some remaining probability, an oracle
is called, representing the second-level urn. This oracle chooses an
existing state in proportion to how many times the oracle previously
chose that state, regardless of the state transition involved, or
chooses a previously unvisited state. The original HDP-HMM provides
an interpretation of this urn model in terms of an underlying
collection of linked random probability measures, however, without
the self-transition parameter. In addition to the conceptual clarity
provided by the random measure formalism, the HDP-HMM has the practical
advantage that it makes it possible to use standard MCMC algorithms for
posterior inference; working within the urn model formulation,
\citet{Beal02} needed to resort to a heuristic approximation to
a Gibbs sampler. The sticky HDP-HMM, an early version of which
was presented in \citet{FoxICML08}, restores the self-transition
parameter of the infinite HMM to this class of models, doing so
in a way that integrates with a full Bayesian nonparametric
specification.

As with the DP, this specification in terms of
random measures yields various interesting characterizations of
marginal probabilities. In particular, as described in the Supplemental
Material~[\citet{SuppA}], the partitioning structure induced by
the sticky HDP-HMM
has an interpretation as an extension of the Chinese restaurant
franchise (CRF) which we refer to as a~\textit{CRF with loyal
customers}. Here, each
restaurant in the franchise has a specialty dish with the same index
as that of the restaurant. Although this dish is served elsewhere,
it is more popular in the dish's namesake restaurant. Recall that while
customers in the CRF of the HDP are
pre-partitioned into restaurants based on the fixed group
assignments, in the HDP-HMM the value of the state $z_{t}$
determines the group assignment (and thus restaurant) of customer
$y_{t+1}$. The increased popularity of the
house specialty dish (determined by the sticky parameter $\kappa$)
implies that children are more likely to eat in
the same restaurant as their parent ($z_t = z_{t-1} = j$) and, in turn,
more likely to eat
the restaurant's specialty dish ($z_{t+1}=j$). This develops family
loyalty to a
given restaurant in the franchise. However, if the parent chooses a
dish other than the house specialty ($z_t = k$, $k \neq j$), the child
will then go to the
restaurant where this dish is the specialty and will in turn be more
likely to eat this dish, too. One might say that for the sticky
HDP-HMM, children have similar taste buds to their parents and will
always go to the restaurant that prepares their parent's dish best.
Often, this keeps many generations eating in the same restaurant.

Throughout the remainder of the paper, we use the following
notational conventions. Given a random sequence
$\{x_1,x_2,\dots,x_T\}$, we use the shorthand $x_{1:t}$ to denote
the sequence $\{x_1,x_2,\dots,x_t\}$ and $x_{\backslash t}$ to
denote the set $\{x_1,\dots,x_{t-1},x_{t+1},\dots,x_T\}$. Also, for
random variables with double subindi\-ces, such as $x_{a_1a_2}$, we
will use $\mathbf{x}$ to denote the entire set of such random variables,
$\{x_{a_1a_2}, \forall a_1, \forall a_2\}$, and the shorthand
notation $x_{a_1\cdot} = \sum_{a_2} x_{a_1a_2}$, $x_{\cdot a_2} =
\sum_{a_1} x_{a_1a_2}$ and $x_{\cdot\cdot} = \sum_{a_1}\sum_{a_2}
x_{a_1a_2}$.

%s5.1 ###
\subsection{Sampling via direct assignments}
\label{sec:revisedHDPHMMinference}

In this section we present an inference algorithm for the sticky
HDP-HMM of Section~\ref{sec:stickyHDPHMM} and Figure~\ref{fig:HDPHMM}(a)
that is a modified version of the direct assignment
Rao-Blackwellized Gibbs sampler of \citet{Teh06}. This sampler
circumvents the complicated bookkeeping of the CRF by sampling
indicator random variables directly. The resulting sticky HDP-HMM
direct assignment Gibbs sampler is outlined in Algorithm~1 of the
Supplementary Material [\citet{SuppA}], which also contains the
full derivations of
this sampler.

The basic idea is that we marginalize over the infinite set of
state-specific transition distributions $\pi_k$ and parameters
$\theta _k$, and sequentially sample the state $z_t$ given all
other state assignments $z_{\backslash t}$, the observations
$y_{1:T}$, and the global transition distribution $\beta$. A variant
of the Chinese restaurant process gives us the prior probability of
an assignment of $z_t$ to a value $k$ based on how many times we
have seen other transitions from the previous state value~$z_{t-1}$
to $k$ and $k$ to the next state value $z_{t+1}$. % We denote the
%number of transitions from $j$ to $k$ in $z_{1:T}$ by $n_{jk}$.
As derived in the Supplementary Material [\citet{SuppA}], this conditional
distribution is dependent upon whether either or both of the
transitions $z_{t-1}$ to $k$ and~$k$ to $z_{t+1}$ correspond to a
self-transition, most strongly when $\kappa>0$. The prior
probability of an assignment of $z_t$ to state $k$ is then weighted
by the likelihood of the observation $y_t$ given all other
observations assigned to state~$k$.

Given a sample of the state sequence $z_{1:T}$, we can represent the
posterior distribution of the global transition distribution $\beta$
via a set of auxiliary random variables $\bar{m}_{jk}$, $m_{jk}$
and $w_{jt}$, which correspond to the $j${th} restaurant-specific
set of table counts associated with the CRF with loyal customers
described in the Supplemental Material [\citet{SuppA}].
%for each considered dish and served dish, and
%override variables of the CRF with loyal customers, respectively.
The Gibbs sampler iterates between sequential sampling of the state
$z_t$ for each individual value of $t$ given $\beta$ and
$z_{\backslash t}$; sampling of the auxiliary variables
$\bar{m}_{jk}$, $m_{jk}$ and~$w_{jt}$ given $z_{1:T}$ and $\beta$;
and sampling of $\beta$ given these auxiliary variables.

The direct assignment sampler is initialized by sampling the
hyperparame\-ters and $\beta$ from their respective priors and then
sequentially sampling each~$z_t$ as if the associated $y_t$ was the
last observation. That is, we first sample~$z_1$ given $y_1$, $\beta$,
and the hyperparameters. We then sample $z_2$ given $z_1$, $y_{1:2}$,~%
$\beta$, and the hyperparameters, and so on. Based on the resulting
sample of $z_{1:T}$, we resample~$\beta$ and the hyperparameters. From
then on, the sampler continues with the normal procedure of
conditioning on $z_{\backslash t}$ when resampling~$z_t$.
\subsection{Blocked sampling of state sequences} \label{sec:blockedz}
\begin{figure}%[t]

\includegraphics{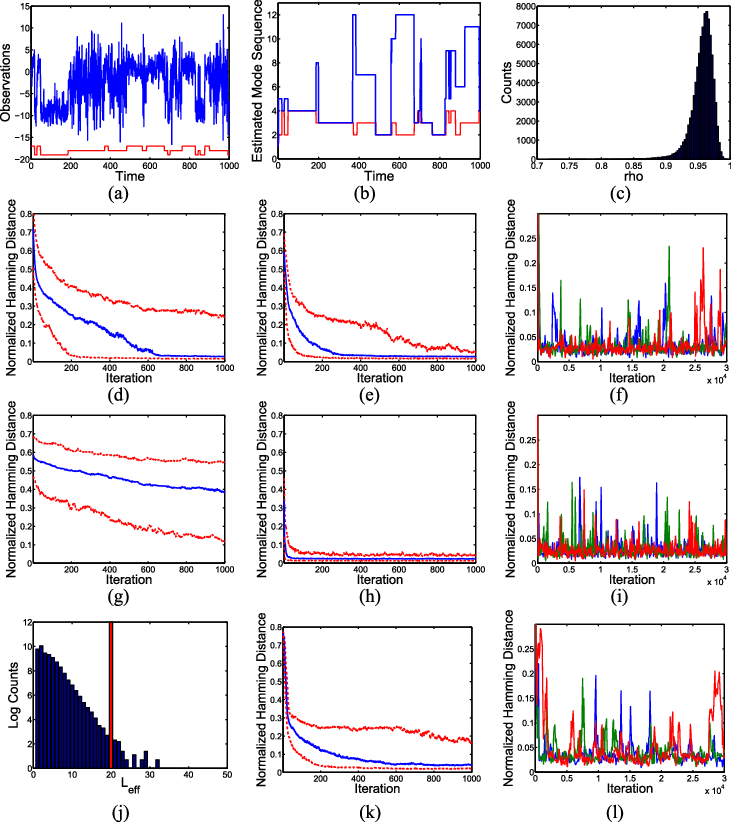}

\caption{\textup{(a)} Observation sequence (blue) and true state
sequence (red) for a three-state HMM with state persistence.
\textup{(b)} Example of the sticky HDP-HMM direct assignment Gibbs sampler
splitting temporally separated examples of the same true state
(red) into multiple estimated states (blue) at Gibbs iteration
{1000}. \textup{(c)} Histogram of the inferred self-transition proportion
parameter, $\rho$, for the sticky HDP-HMM blocked sampler.
For the original HDP-HMM, the median (solid blue) and
{10}{th} and {90}{th} quantiles (dashed red) of Hamming distance
between the true and estimated state sequences over the first {1000}
Gibbs samples from {200} chains are shown for the \textup{(d)} direct assignment
sampler, and \textup{(e)} blocked sampler. \textup{(f)} Hamming distance over {30,000}
Gibbs samples from three chains of the original HDP-HMM blocked sampler.
\textup{(g)}--\textup{(i)} Analogous plots to \textup{(d)} and \textup{(f)} for the sticky
HDP-HMM. \textup{(k)} and \textup{(l)} Plots analogous to \textup{(e)} and \textup{(f)} for
 a nonsticky HDP-HMM using beam sampling. \textup{(j)} A
histogram of the effective beam sampler
truncation level, $L_{\mathrm{eff}}$, over the {30,000} Gibbs iterations from
the three chains (blue) compared to the fixed truncation level,
$L=20$, used in the truncated sticky HDP-HMM blocked sampler results
(red).}\label{fig:state_persistence}
\end{figure}

%
% \centering
% \begin{tabular}{ccc}
% \hspace{-0.1in}\includegraphics[height =
%1.6in]{JournalFigs/obs_1D_3mode}
% %& \includegraphics[height=1.6in]{JournalFigs/directEstSeq_9}
% & \includegraphics[height=1.6in]{JournalFigs/directEstSeq_51}
% %& \includegraphics[height=1.15in]{JournalFigs/directEstSeq_180}\\
% \hspace{-0.1in}{\small(a)} & {\small(b)} \precap%& (c)% & (d)
% \end{tabular}
% \caption{Plots showing the sequential Gibbs sampler splitting
%temporally
%separated examples of the same true state into multiple estimated
%states. (a) Observation sequence. (b) Example of the
%estimated HMM state sequence (blue) at Gibbs iteration 1000
%overlayed on the true HMM state sequence (red). Here, we
%see that a single true state is divided into multiple estimated
%states, each with high probability of self-transition.}
%
The HDP-HMM sequential, direct assignment sampler of
Section~\ref{sec:revisedHDPHMMinference} can exhibit slow mixing rates
since global state sequence changes are forced to occur coordinate
by coordinate. This phenomenon is explored in \citet{Scott02} for
the finite HMM. Although the sticky HDP-HMM reduces the posterior
uncertainty caused by fast state-switching explanations of the data,
the self-transition bias can cause two continuous and temporally
separated sets of observations of a given state to be grouped into
two states. See Figure~\ref{fig:state_persistence}(b) for an example.
If this occurs, the high probability of self-transition makes it
challenging for the sequential sampler to group those two examples
into a single state.

%f6 ###

%A variant of the HMM forward--backward procedure \citet{Rabiner89}
%allows us to harness the Markovian structure and jointly sample the
%state sequence $z_{1:T}$ given the observations $y_{1:T}$,
%transition probabilities $\pi_k$, and parameters $\uniqueTheta_k$.
We thus propose using a variant of the HMM
forward--backward procedure [\citet{Rabiner89}] to harness the
Markovian structure and jointly sample the state sequence $z_{1:T}$
given the observations $y_{1:T}$, transition probabilities~$\pi_k$,
and parameters $\theta _k$. There are two main mechanisms for
sampling in an uncollapsed HDP model (i.e., one that instantiates the
parameters $\pi_k$ and $\theta _k$): one is to employ slice
sampling while the other is to consider a~truncated approximation to
the HDP. For the HDP-HMM, a slice sampler, referred to as \textit{beam
sampling}, was recently developed~[\citet{VanGael08}]. This
sampler harnesses the efficiencies of the forward--backward algorithm
without having to fix a truncation level for the HDP. However, as we
elaborate upon in Section~\ref{sec:GaussResults}, this sampler suffers
from slower mixing rates than the block sampler we propose, which
utilizes a fixed-order truncation of the HDP-HMM. Although a fixed
truncation reduces our model to a parametric Bayesian HMM, the specific
hierarchical prior induced by a truncation of the fully nonparametric
HDP significantly improves upon classical parametric Bayesian HMMs.
Specifically, a fixed degree $L$ truncation encourages each transition
distribution to be sparse over the set of $L$ possible HMM states, and
simultaneously encourages transitions from different states to have
similar sparsity structures. That is, the truncated HDP prior leads to
a \textit{shared} sparse subset of the $L$ possible states. See
Section~\ref{sec:sparseDir} for a comparison with standard parametric modeling.

There are multiple methods of approximating the countably infinite
transition distributions via truncations. One approach is to terminate
the stick-breaking construction after some portion of the stick has
already been broken and assign the remaining weight to a single
component. This approximation is referred to as the \textit{truncated
Dirichlet process}. Another method is to consider the \textit{degree $L$
weak limit approximation} to the DP [\citet{Ishwaran02-2}],
%
%e5.2 ###
\begin{eqnarray}
\operatorname{GEM}_L(\alpha) \triangleq\operatorname{Dir}(\alpha/L,\dots,\alpha/L),
\end{eqnarray}
where $L$ is a number that exceeds the total number of expected HMM
states. Both of these approximations, which are presented in
\citeauthor{Ishwaran00} (\citeyear{Ishwaran00,Ishwaran02-2}), encourage the learning of models
with fewer than $L$ components while allowing the generation of new
components, upper bounded by~$L$, as new data are observed. We
choose to use the second approximation because of its simplicity and
computational efficiency. The two choices of approximations are
compared in \citet{Kurihara07}, and little to no practical
differences are found. Using a weak limit approximation to the
Dirichlet process prior on
$\beta$ (i.e., employing a finite Dirichlet prior) induces a finite
Dirichlet prior on $\pi_j$:
%
%e5.4 ###
%e5.3 ###
\begin{eqnarray}\label{eqn:truncatedHDPdirichlets}
\beta| \gamma&\sim&\operatorname{Dir}(\gamma/L,\dots,\gamma/L),
\\
\pi_j| \alpha,\beta &\sim&
\operatorname{Dir}(\alpha\beta_1,\dots,\alpha\beta_L).
\end{eqnarray}
As $L\rightarrow\infty$, this model converges in distribution to
the HDP mixture model [\citet{Teh06}].

The Gibbs sampler using blocked resampling of $z_{1:T}$ is derived
in the Supplementary Material [\citet{SuppA}]; an outline of the
resulting algorithm
is also presented (see Algorithm 3). A similar sampler has been used
for inference in HDP hidden Markov trees [\citet{Kivinen07}].
However, this work did not consider the complications introduced by
multimodal emissions, which we explore in
Section~\ref{sec:temperedHDPHMMDP}.

The blocked sampler is initialized by drawing $L$ parameters
$\theta _k$ from the base measure, $\beta$ from its
$L$-dimensional symmetric Dirichlet prior, and the~$L$ transition
distributions $\pi_k$ from the induced $L$-dimensional Dirichlet prior
specified in equation~\eqref{eqn:truncatedHDPdirichlets}. The
hyperparameters are also drawn from the prior. Based on the sampled
parameters and transition distributions, one can block sample $z_{1:T}$
and proceed as in Algorithm 3 of the Supplementary Material [Fox et al.
(\citeyear{SuppA})].
\subsection{Hyperparameters} \label{sec:hypers}
We treat the hyperparameters in the sticky HDP-HMM as unknown
quantities and perform full Bayesian inference over these
quantities. This emphasizes the role of the data in determining the
number of occupied states and the degree of self-transition bias.
Our derivation of sampling updates for the hyperparameters of the
sticky HDP-HMM is presented in the Supplementary Material~[\citet
{SuppA}]; it
roughly follows that of the original HDP-HMM~[\citet{Teh06}]. A key
step which simplifies our inference procedure is to note that since
we have the deterministic relationships\vspace*{-2pt}
\begin{eqnarray}
\alpha&=& (1-\rho)(\alpha+ \kappa),\nonumber
\\[-9pt]\\[-9pt]
\kappa&=& \rho(\alpha+ \kappa),\nonumber\vspace*{-2pt}
\end{eqnarray}
we can treat $\rho$ and $\alpha+\kappa$ as our hyperparameters and
sample these values instead of sampling $\alpha$ and $\kappa$
directly.

%s6 ###
\section{Experiments with synthetic data}
\label{sec:results1}
In this section we explore the perfor\-mance of the sticky HDP-HMM
relative to the original model (i.e., the model with $\kappa= 0$)
in a series of experiments with synthetic data. We judge
perfor\-mance according to two metrics: our ability to accurately
segment the data according to the underlying state sequence, and the
predictive likelihood of held-out data under the inferred model. We
additionally assess the improvements in mixing rate achieved by
using the blocked sampler of Section~\ref{sec:blockedz}.

%s6.1 ###
\subsection{Gaussian emissions}
\label{sec:GaussResults}
We begin our analysis of the sticky HDP-HMM performance by examining
a set of simulated data generated from an HMM with Gaussian
emissions. The first data set is generated from an HMM with a high
probability of self-transition. Here, we aim to show that the
original HDP-HMM inadequately captures state persistence. The second
data set is from an HMM with a high probability of leaving the
current state. In this scenario, our goal is to demonstrate that the
sticky HDP-HMM is still able to capture rapid dynamics by inferring
a small probability of self-transition.

For all of the experiments with simulated data, we used weakly
informative hyperpriors. We placed a $\operatorname{Gamma}(1,0.01)$ prior on
the concentration parameters $\gamma$ and $(\alpha+\kappa)$. The
self-transition proportion parameter $\rho$ was given a
$\operatorname{Beta}(10,1)$ prior. The parameters of the base measure were
set from the data, as will be described for each scenario.

\subsubsection*{State persistence}
The data for the high persistence case were generated from a
three-state HMM with a 0.98 probability of self-transition and equal
probability of transitions to the other two states. The observation
and true state sequences for the state persistence scenario are
shown in Figure~\ref{fig:state_persistence}(a). We placed a normal
inverse-Wishart prior on the space of mean and variance
parameters and set the hyperparameters as follows: %pseudocounts $\zeta
%=
%0.01$, mean $\vartheta= \frac{1}{T}\sum_{t=1}^T y_{t}$, degrees of
%freedom $\nu= 3$, and scale matrix $S = \frac{1}{T-1}\sum_{t=1}^T
%(y_t-\vartheta)(y_t-\vartheta)^T$
0.01 pseudocounts, mean equal to the empirical mean, three degrees
of freedom, and scale matrix equal to 0.75 times the empirical
variance. We used this conjugate base measure~so that we may
directly compare the performance of the blocked and direct
as\-signment samplers. % of Algorithms 1 and 3.
For the blocked sampler, we used a truncation level of $L=20$.

In Figure~\ref{fig:state_persistence}(d)--(h), we plot the $10${th},
$50${th} and $90${th} quantiles of the Hamming distance between
the true and estimated state sequences over the 1000 Gibbs
iterations using the direct assignment and blocked samplers on the
sticky\vadjust{\goodbreak} and original HDP-HMM models. To calculate the Hamming
distance, we used the Munkres algorithm [\citet{Munkres57}] to map
the randomly chosen indices of the estimated state sequence to the
set of indices that maximize the overlap with the true sequence.

From these plots, we see that the burn-in rate of the blocked
sampler using the sticky HDP-HMM is significantly faster than that
of any other sampler-model combination. As expected, the sticky
HDP-HMM with the sequential, direct assignment sampler
%(Algorithm~\ref{alg:direct})
gets stuck in state sequence assignments from which it is hard to
move away, as conveyed by the flatness of the Hamming error versus
iteration number plot in Figure~\ref{fig:state_persistence}(g). For
example, the estimated state sequence of
Figure~\ref{fig:state_persistence}(b) might have similar parameters
associated with states 3, 7, 10 and 11 so that the likelihood is in
essence the same as if these states were grouped, but this sequence
has a large error in terms of Hamming distance and it would take
many iterations to move away from this assignment. Incorporating the
blocked sampler with the original HDP-HMM improves the Hamming
distance performance relative to the sequential, direct assignment
sampler for both the original and sticky HDP-HMM; however, the
burn-in rate is still substantially slower than that of the blocked
sampler on the sticky model. % (Algorithm~\ref{alg:blocked}).

As discussed earlier, a \textit{beam sampling} algorithm [\citet
{VanGael08}] has
been proposed which adapts slice sampling methods [\citet
{Robert}] to
the HDP-HMM. This approach uses a set of auxiliary slice variables,
one for each observation, to effectively truncate the number of
state transitions that must be considered at every Gibbs sampling
iteration. Dynamic programming methods can then be used to jointly
resample state assignments. The beam sampler was inspired by a
related approach for DP mixture models [\citet{Walker07}], which is
conceptually similar to retrospective sampling methods
[\citet{Papaspiliopoulos08}]. In comparison to our fixed-order,
weak-limit truncation of the HDP-HMM, the beam sampler provides an
asymptotically exact algorithm. However, the beam sampler can be
slow to mix relative to our blocked sampler on the fixed, truncated
model (see Figure~\ref{fig:state_persistence} for an example
comparison). The issue
is that in order to consider a transition which has low prior
probability, one needs a correspondingly rare slice variable sample
at that time. Thus, even if the likelihood cues are strong, to be
able to consider state sequences with several low-prior-probability
transitions, one needs to wait for several \textit{rare events} to
occur when drawing slice variables. By considering the full,
exponentially large set of paths in the truncated state space, we
avoid this problem. Of course, the trade-off between the
computational cost of the blocked sampler on the fixed, truncated
model ($O(TL^2)$) and the slower mixing rate of the beam sampler
yields an application-dependent sampler choice.

The Hamming distance plots of Figure~\ref{fig:state_persistence}(k)
and (l), when
compared to those of Figure~\ref{fig:state_persistence}(e) and (f),
depict the
substantially slower mixing rate of the beam sampler compared to the
blocked sampler (both using a nonsticky HDP-HMM). However, the
theoretical computational benefit of
the beam sampler can be seen in Figure~\ref{fig:state_persistence}(j).
In this
plot, we present a histogram~of the effective truncation level,
$L_{\mathrm{eff}}$, used over the 30,000 Gibbs iterations on three chains. We
computed this effective truncation level by summing over the number
of state transitions considered during a full sweep of\vadjust{\eject} sampling~%
$z_{1:T}$ and then dividing this number by the length of the
data set, $T$, and taking the square root. Finally, on a more technical note,
our fixed, truncated model allows for more vectorization of the code
than the beam sampler. Thus, in practice, the difference in
computation time between the samplers is significantly less than the
$O(L^2/L^2_{\mathrm{eff}})$ factor obtained by counting state transitions.

From this point onward, we present results only from blocked
sampling since we have seen the clear advantages of this method over
the sequential, direct assignment sampler.\vspace*{-3pt}

\subsubsection*{Fast state-switching}
In order to warrant the general use of the sticky model, one would
like to know that the sticky parameter incorporated in the model does not
preclude learning models with fast dynamics. To this end, we
explored the performance of the sticky HDP-HMM on data generated
from a model with a high probability of switching between states.
Specifically, we generated observations from a four-state HMM with
the following transition probability matrix:\vspace*{-5pt}
%
%e6.1 ###
\begin{eqnarray}
\left[\matrix{
0.4 & 0.4 & 0.1 & 0.1\cr
0.4 & 0.4 & 0.1 & 0.1\cr
0.1 & 0.1 & 0.4 & 0.4\cr
0.1 & 0.1 & 0.4 & 0.4
}\right]
.
\end{eqnarray}
We once again used a truncation level $L=20$. Since we are
restricting ourselves to the blocked Gibbs sampler, it is no longer
necessary to use a conjugate base measure. Instead we placed an
independent Gaussian prior on the mean parameter and an
inverse-Wishart prior on the variance parameter. For the Gaussian
prior, we set the mean and variance hyperparameters to be equal to
the empirical mean and variance of the entire data set. The
inverse-Wishart hyperparameters were set such that the expected
variance is equal to 0.75 times that of the entire data set, with
three degrees of freedom.

%f7 ###
\begin{figure*}[t]

\includegraphics{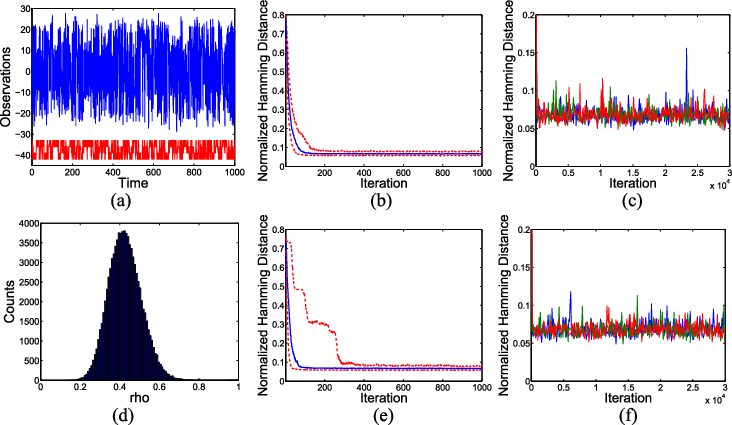}%
\vspace*{-5pt}
\caption{\textup{(a)} Observation sequence (blue) and true state
sequence (red) for a four-state HMM with fast state switching.
For the original HDP-HMM using a blocked Gibbs sampler:
\textup{(b)} the median (solid blue) and {10}{th} and {90}{th}
quantiles (dashed red) of Hamming distance between the true and
estimated state sequences over the first {1000} Gibbs samples from
{200} chains, and {(c)} Hamming distance over {30,000} Gibbs samples from
three chains. \textup{(d)}~Histogram of the inferred self-transition
parameter, $\rho$, for the sticky HDP-HMM blocked sampler. \textup{(e)} and \textup{(f)}
Analogous plots to \textup{(b)} and \textup{(c)} for the sticky HDP-HMM.}
\label{fig:fast_switching}\vspace*{-6pt}
\end{figure*}
%
%%
% \centering
% \begin{tabular}{cc}
% \includegraphics[height =
%1.6in]{JournalFigs/hamming_dist_fastswitching_kappa_burnin}
% & \includegraphics[height =
%1.6in]{JournalFigs/hamming_dist_fastswitching_nokappa_burnin}
% {\small(a)} & {\small(b)}
% \end{tabular}
% \caption{For the observation sequence of Fig.~
% the median (solid blue) and $10^{th}$ and $90^{th}$
%quantiles (dashed red) of Hamming distance between the true and
%estimated state sequences over the first 1,000 Gibbs samples from
%200 chains are shown for the (a) sticky HDP-HMM and (b) original
%HDP-HMM using the blocked sampler.}
% \label{fig:fast_switching2}
%%

The results depicted in Figure~\ref{fig:fast_switching}
%Fig.~\ref{fig:fast_switching2}
confirm that by inferring a small probability of self-transition,
the sticky HDP-HMM is indeed able to capture fast HMM dynamics, and
just as quickly as the original HDP-HMM (although with higher
variability). Specifically, we see that the histogram of the
self-transition proportion parameter $\rho$ for this data set [see
Figure~\ref{fig:fast_switching}(d)] is centered around a value close
to the true probability of self-transition, which is substantially
lower than the mean value of this parameter on the data with high
persistence [Figure~\ref{fig:state_persistence}(c)].\vspace*{-3pt}

%s6.2 ###
\subsection{Multinomial emissions}
\label{sec:MultResults}
The difference in modeling power, rather than simply burn-in rate,
between the sticky and original HDP-HMM is more pronounced when we
consider multinomial emissions. This is because the multinomial
observations are embedded in a discrete topological space in which
there is no concept of similarity between nonidentical
observation
values. In contrast, Gaussian emissions have a continuous range of
values in~$\mathbb{R}^n$\vadjust{\eject} with a clear notion of \textit{closeness}
between observations under the Lebesgue measure, aiding in grouping
observations under a single HMM state's Gaussian emission
distribution, even in the absence of a self-transition bias.

\begin{figure*}[t]

\includegraphics{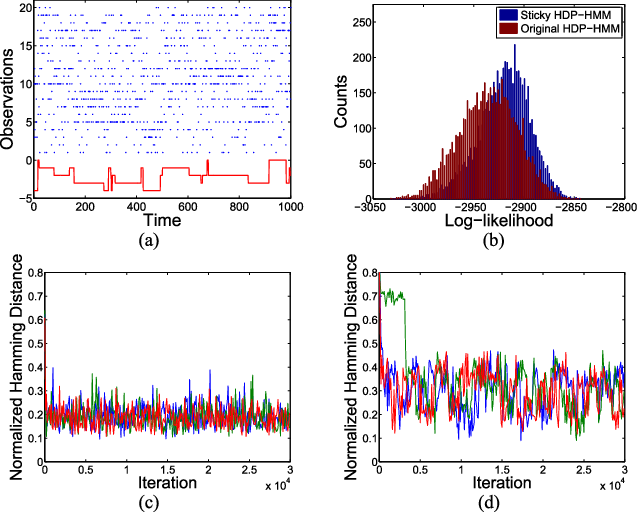}
\vspace*{-5pt}
\caption{\textup{(a)} Observation sequence (blue) and true state
sequence (red) for a five-state HMM with multinomial observations. \textup{(b)}
Histogram of the predictive probability of test sequences using the
inferred parameters sampled every {100}{th} iteration from Gibbs
iterations {10,000--30,000} for the sticky and original HDP-HMM. The
Hamming distances over {30,000} Gibbs samples from three chains are
shown for the \textup{(c)} sticky HDP-HMM and \textup{(d)} original HDP-HMM.}
\label{fig:multinomial}\vspace*{-6pt}
\end{figure*}

To demonstrate the increased posterior uncertainty with discrete
observations, we generated data from a five-state HMM with
multinomial emissions with a 0.98 probability of self-transition and
equal probability of transitions to the other four states. The
vocabulary, or range of possible observation values, was set to 20.
The observation and true state sequences are shown in
Figure~\ref{fig:multinomial}(a). We placed a symmetric Dirichlet
prior on the parameters of the multinomial distribution, with the
Dirichlet hyperparameters equal to 2 [i.e.,
$\operatorname{Dir}(2,\dots,2)]$.

%f8 ###

From Figure~\ref{fig:multinomial}, we see that even after burn-in,
many fast-switching state sequences have significant posterior
probability under the nonsticky model, leading to sweeps through
regions of larger Hamming distance error. A qualitative plot of one
such inferred sequence after 30,000 Gibbs iterations is shown in
Figure~\ref{fig:rapid_dynamics}(c). Such sequences have negligible
posterior probability under the sticky HDP-HMM formulation.

In some applications, such as the speaker diarization problem that
is explored in Section~\ref{sec:SpeakerDiarization}, one cares about
the inferred segmentation of the data into a set of state labels. In
this case, the advantage of incorporating the sticky parameter is
clear. However, it is often the case that the metric of interest is
the predictive power of the fitted model, not the accuracy of the
inferred state sequence. To study performance under this metric, we
simulated 10 test sequences using the same parameters that generated
the training sequence. We then computed the likelihood of each of
the test sequences under the set of parameters inferred at every
$100${th} Gibbs iteration from iterations 10,000--30,000. This
likelihood was computed by running the forward--backward algorithm of
\citet{Rabiner89}. We plot these results as a~histogram in
Figure~\ref{fig:multinomial}(b). From this plot, we see that the
fragmentation of data into redundant HMM states can also degrade the
predictive performance of the inferred model. Thus, the sticky
parameter plays an important role in the Bayesian nonparametric
learning of HMMs even in terms of model averaging.\looseness=-1\vspace*{-3pt}

%s6.3 ###
\subsection{Comparison to independent sparse Dirichlet prior}
\label{sec:sparseDir}
\begin{figure*}[t]

\includegraphics{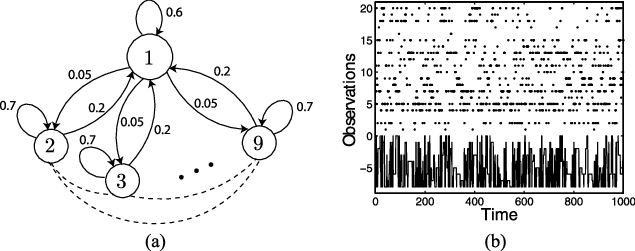}
\vspace*{-5pt}
\caption{\textup{(a)} State transition diagram for a nine-state HMM with one
main state (labeled 1) and eight sub-states (labeled {2--9}). All
states have a significant probability of self-transition. From the main
state, all other states are equally likely. From a sub-state, the
most likely nonself-transition is a transition back to the main
state. However, all sub-states have a~small probability of
transitioning to another sub-state, as indicated by the dashed
arcs. \textup{(b)}~Observation sequence (top) and true state sequence
(bottom) generated by the nine-state HMM with multinomial observations.}
\label{fig:sparseDirHMM}\vspace*{-6pt}
\end{figure*}

\begin{figure*}[t]

\includegraphics{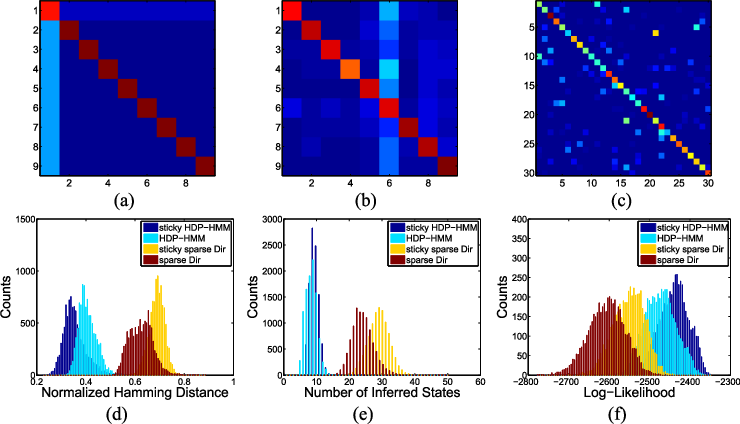}%
\vspace*{-5pt}
\caption{\textup{(a)} The true transition probability matrix (TPM)
associated with the state transition diagram of Figures~\protect\ref{fig:sparseDirHMM}.
\textup{(b)} and \textup{(c)} The inferred TPM at the {30,000}th Gibbs iteration for the
sticky HDP-HMM and sticky sparse Dirichlet model, respectively, only
examining those states with more than {1}\% of the assignments. For
the HDP-HMM and sparse Dirichlet model, with and without the sticky
parameter, we plot: \textup{(d)} the Hamming distance error over {10,000} Gibbs
iterations, \textup{(e)} the inferred number of states with more than {1}\% of
the assignments, and \textup{(f)} the predictive probability of test
sequences using the inferred parameters sampled every {100}{th}
iteration from Gibbs iterations {5000--10,000}. }
\label{fig:sparseDir}\vspace*{-5pt}
\end{figure*}

We have alluded to the fact that the \textit{shared} sparsity of the
HDP-HMM induced by $\beta$ is essential for inferring sparse
representations of the data. Although this is clear from the
perspective of the prior model, or, equivalently, the generative
process, it is not immediately obvious how much this hierarchical
Bayesian\vadjust{\eject} constraint helps us in posterior inference. Once we are in
the realm of considering a fixed, truncated approximation to the
HDP-HMM, one might propose an alternate model in which we simply
place a sparse Dirichlet prior,
$\operatorname{Dir}(\alpha/L,\dots,\alpha/L)$ with $\alpha/L < 1$,
independently on each row of the transition matrix. This is
equivalent to setting $\beta= [1/L,\dots,1/L]$ in the truncated
HDP-HMM, which can also be achieved by letting the hyperparameter
$\gamma$ tend to infinity. Indeed, when the data do not exhibit
shared sparsity or when the likelihood cues are sufficiently strong,
the independent sparse Dirichlet prior model can perform as well as
the truncated HDP-HMM. However, in scenarios such as the one
depicted in Figure~\ref{fig:sparseDirHMM}, we see substantial
differences in performance by considering the HDP-HMM, as well as
the inclusion of the sticky parameter. We explored the relative
performance of the HDP-HMM and sparse Dirichlet prior model, with
and without the sticky parameter, on such a Markov model with
multinomial emissions on a vocabulary of size 20. We placed a
$\operatorname{Dir}(0.1,\dots,0.1)$ prior on the parameters of the
multinomial distribution. For the sparse Dirichlet prior model, we
assumed a state space of size 50, which is the same as the
truncation level we chose for the HDP-HMM (i.e., $L=50$). The
results are presented in Figure~\ref{fig:sparseDir}. From these plots,
we see that the hierarchical Bayesian approach of the HDP-HMM does,
in fact, improve the fitting of a model with shared sparsity. The
HDP-HMM consistently infers fewer HMM states and more representative
model parameters. As a result, the HDP-HMM has higher predictive
likelihood on test data, with an additional benefit gained from
using the sticky parameter.

%f9 ###

%f10 ###

Note that the results of Figure~\ref{fig:sparseDir}(f) also motivate
the use of the sticky parameter in the more classical setting of a
finite HMM with a standard Dirichlet sparsity prior. A motivating
example of the use of sparse Dirichlet priors for finite HMMs is
presented in~\citet{Johnson07}.

%s7 ###
\section{Multimodal emission densities}
\label{sec:temperedHDPHMMDP}
In many application domains, the data associated with each hidden
state may have a complex, multimodal distribution. We propose to
model such emission distributions nonparametrically, using a DP
mixture of Gaussians. This formulation is related to the nested
DP~[\citet{Rodriguez08}], which uses a Dirichlet process to partition
data into groups, and then models each group via a Dirichlet process
mixture. The bias toward self-transitions allows us to distinguish
between the underlying HDP-HMM states. If the model were free to
both rapidly switch between HDP-HMM states and associate multiple
Gaussians per state, there would be considerable posterior
uncertainty. Thus, it is only with the sticky HDP-HMM that we can
effectively fit such models.

We augment the HDP-HMM state $z_t$ with a term $s_t$ indexing the
mixture component of the $z_t${th} emission density. For each
HDP-HMM state, there is a unique stick-breaking measure $\psi_k \sim
\operatorname{GEM}(\sigma)$ defining the mixture weights of the $k${th}
emission density so that $s_t \sim\psi_{z_t}$. Given the augmented
state $(z_t,s_t)$, the observation $y_t$ is generated by the
Gaussian component with parameter $\theta _{z_t,s_t}$. Note
that both the HDP-HMM state index and mixture component index are
allowed to take values in a countably infinite set. See
Figure~\ref{fig:HDPHMM}(b).

%s7.1 ###
\subsection{Direct assignment sampler}
\label{sec:HDPHMMDPdirectsampler}
Many of the steps of the direct assignment sampler for the sticky
HDP-HMM with DP emissions remain the same as for the regular sticky
HDP-HMM. Specifically, the sampling of the global transition
distribution $\beta$, the table counts $m_{jk}$ and $\bar{m}_{jk}$,
and the override variables $w_{jt}$ are unchanged. The difference
arises in how we sample the augmented state $(z_t,s_t)$.

The joint distribution on the augmented state, having marginalized
the transition distributions $\pi_k$ and emission mixture weights
$\psi_k$, is given by
\begin{eqnarray}
&&p(z_t=k,s_t=j |  z_{\backslash t},s_{\backslash
t},y_{1:T},\beta,\alpha,\sigma,\kappa,\lambda)\nonumber
\\[-1pt]
&&\qquad = p(s_t=j|
z_t=k,z_{\backslash t},s_{\backslash
t},y_{1:T},\sigma,\lambda),\nonumber
\\[-1pt]
&&\hspace*{20pt}p(z_t=k|  z_{\backslash t},s_{\backslash t},
y_{1:T},\beta,\alpha,\kappa,\lambda).\nonumber
\end{eqnarray}
We then block-sample $(z_t,s_t)$ by first sampling $z_t$, followed
by $s_t$ conditioned on the sampled value of $z_t$. The term
$p(s_t=j|  z_t=k,z_{\backslash t},s_{\backslash
t},y_{1:T},\sigma,\lambda)$ relies on how many observations are
currently assigned to the $j${th} mixture component of state $k$.
%,which we denote by $n'_{kj}$.
These conditional distributions are derived in the Supplementary
Material~[\citet{SuppA}], which also contains an outline of the resulting Gibbs sampler
 in Algorithm 2.
\subsection{Blocked sampler}
To implement blocked resampling of $(z_{1:T},s_{1:T})$, we use weak
limit approximations to both the HDP-HMM and DP emissions,
approximated to levels $L$ and $L'$, respectively. The posterior
distributions for $\beta$ and $\pi_k$ remain unchanged from the
sticky HDP-HMM; that of $\psi_k$ is given by
%
%e7.1 ###
\begin{eqnarray}
\psi_k|  z_{1:T},s_{1:T},\sigma\sim\operatorname{Dir}(\sigma/L' +
n'_{k1},\dots,\sigma/L'+n'_{kL'}),
\end{eqnarray}
where $n'_{k\ell}$ is the number of $s_t$ taking a value $\ell$ when $z_t=k$.
(i.e., the number of observations assigned to the $k${th} state's
$\ell${th} mixture
component). The procedure for sampling the augmented state $(z_{1:T},s_{1:T})$
is derived in the Supplementary Material [see Algorithm 4,~\citet
{SuppA}]. %%
\subsection{Assessing the multimodal emissions model}
\label{sec:MoGResults}
\begin{figure*}[t]

\includegraphics{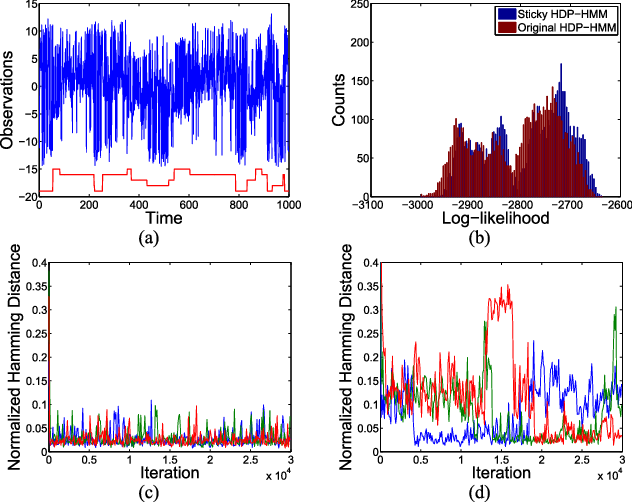}
\vspace*{-5pt}
\caption{\textup{(a)} Observation sequence (blue) and true state
sequence (red) for a five-state HMM with mixture-of-Gaussian observations.
\textup{(b)} Histogram of the predictive probability of test
sequences using the inferred parameters sampled every {100}{th}
iteration from Gibbs iterations {10,000--30,000} for the sticky and
original HDP-HMM.
The Hamming distance over {30,000} Gibbs samples from three chains are
shown for the \textup{(c)} sticky HDP-HMM and \textup{(d)} original HDP-HMM, both with
DP emissions.}
\label{fig:MoG}%
\vspace*{-6pt}
\end{figure*}

In this section we evaluate the ability of the sticky HDP-HMM to
infer multimodal emission distributions relative to the model
without the sticky parameter. We generated data from a five-state
HMM with mixture-of-Gaussian emissions, where the number of mixture
components for each emission distribution was chosen randomly from a
uniform distribution on $\{1, 2, \ldots, 10\}$. Each component of
the mixture was equally weighted and the probability of
self-transition was set to 0.98, with equal probabilities of
transitions to the other states. The large probability of
self-transition is what disambiguates this process from one with
many more HMM states, each with a single Gaussian emission
distribution. The resulting observation and true state sequences are
shown in Figure~\ref{fig:MoG}(a).

We once again used a nonconjugate base measure and placed a
Gaussian prior on the mean parameter and an independent
inverse-Wishart prior on the variance parameter of each Gaussian
mixture component. The hyperparameters for these distributions were
set from the data in the same manner as in the fast-switching
scenario. Consistent with the sticky HDP-HMM concentration
parameters $\gamma$ and $(\alpha+\kappa)$, we placed a weakly
informative $\operatorname{Gamma}(1,0.01)$ prior on the concentration
parameter $\sigma$ of the DP emissions. All results are for the
blocked sampler with truncation levels $L=L'=20$.

%f11 ###

In Figure~\ref{fig:MoG} we compare the performance of the sticky
HDP-HMM with DP emissions to that of the original HDP-HMM with DP
emissions (i.e., DP emissions, but no bias toward
self-transitions). As with the multinomial observations, when the
distance between observations does not directly factor into the
grouping of observations into HMM states, there is a considerable
amount of posterior uncertainty in the underlying HMM state of the nonsticky model. Even
after 30,000 Gibbs samples, there are still state sequence sample
paths with very rapid dynamics. The result of this fragmentation
into redundant states is a slight reduction in predictive
performance on test sequences, as in the multinomial emission case.
See Figure~\ref{fig:MoG}(b).

%s8 ###
\section{Speaker diarization results}
\label{sec:SpeakerDiarization}
Recall the \textit{speaker diarization} task from Section~\ref
{sec:SpeakerDiarizationTask}, which involves segmenting
audio recordings from the NIST Rich Transcription 2004--2007 database
into speaker-homogeneous regions while
simultaneously identifying the number of speakers. In this section
we present our results on applying the sticky HDP-HMM with DP
emissions to the speaker diarization task.

A minimum speaker duration of 500~ms was set by associating two
preprocessed MFCCs
with each hidden state. We also tied the covariances of within-state
mixture components (i.e., each speaker-specific mixture component
was forced to have identical covariance structure), and used a
nonconjugate prior on the mean and covariance parameters. We placed
a normal prior on the mean parameter with mean equal to the
empirical mean and covariance equal to 0.75 times the empirical
covariance, and an inverse-Wishart prior on the covariance parameter
with 1000 degrees of freedom and expected covariance equal to the
empirical covariance. Our choice of a large degrees of freedom is akin
to an empirical Bayes approach in that it concentrates the mass of the
prior in reasonable regions based on the data. Such an approach is
often helpful in high-dimensional applied problems since our sampler
relies on forming new states (i.e., speakers) based on parameters drawn
from the prior. Issues of exploration in this high-dimensional space
increase the importance of the setting of the base measure. For the
concentration parameters, we placed a
$\operatorname{Gamma}(12,2)$ prior on $\gamma$, a $\operatorname{Gamma}(6,1)$ prior
on $\alpha+\kappa$, and a $\operatorname{Gamma}(1,0.5)$ prior on $\sigma$.
The self-transition parameter $\rho$ was given a
$\operatorname{Beta}(500,5)$ prior. For each of the 21 meetings, we ran 10
chains of the blocked Gibbs sampler %of Algorithm~
for 10,000 iterations for both the original and sticky HDP-HMM with
DP emissions. We used a sticky HDP-HMM truncation level of $L=15$,
where the DP-mixture-of-Gaussians emission distribution associated with
each of these $L$ HMM states was truncated to $L' = 30$ components. Our
choice of $L$ significantly exceeds the typical number of speakers,
which in the NIST database tends to be between 4 and 6. In practice,
our sampler never approached using the full set of possible states and
emission components.

In order to explore the importance of capturing the temporal dynamics,
we also compare our sticky HDP-HMM performance to that of a Dirichlet
process mixture of Gaussians that simply pools together the data from
each meeting, ignoring the time indices associated with the
observations. We considered a truncated Dirichlet process mixture model
with $L=50$ components and a $\operatorname{Gamma}(6,1)$ prior on the
concentration parameter $\gamma$. The base measure was set as in the
sticky HDP-HMM.

For the NIST speaker diarization evaluations, the goal is to produce
a~single segmentation for each meeting. Due to the label-switching
issue (i.e., under our exchangeable prior, labels are arbitrary
entities that do not necessarily remain consistent over Gibbs
iterations), we cannot simply integrate over multiple Gibbs-sampled
state sequences. We propose two solutions to this problem. The
first, which we refer to as the \textit{likelihood metric}, is to simply
choose from a fixed set of Gibbs samples the one
that produces the largest likelihood given the estimated parameters
(marginalizing over state sequences), and then produce the
corresponding Viterbi state sequence. This heuristic, however, is
sensitive to overfitting and will, in general, be biased toward
solutions with more states.

An alternative, and more robust, metric is what we refer to as the
\textit{minimum expected Hamming distance}. We first choose a large
reference set $\mathcal{R}$ of state sequences produced by the Gibbs
sampler and a possibly smaller set of test sequences $\mathcal{T}$.
Then, for each sequence $z^{(i)}$ in the test set $\mathcal{T}$, we
compute the empirical mean Hamming distance between the test
sequence and the sequences in the reference set $\mathcal{R}$; we
denote this empirical mean by $\hat{H}_i$. We then choose the test
sequence $z^{(j^*)}$ that minimizes this expected Hamming distance. That
is,
\begin{eqnarray*}
z^{(j^*)} = \arg\min_{z^{(i)} \in\mathcal{T}} \hat{H}_i.
\end{eqnarray*}
The empirical mean Hamming distance $\hat{H}_i$ is a
\textit{label-invariant loss function} since it does not rely on
labels remaining consistent across samples---we simply compute
\begin{eqnarray*}
\hat{H}_i = \frac{1}{|\mathcal{R}|} \sum_{z^{(j)} \in\mathcal{R}}
\operatorname{Hamm}\bigl(z^{(i)},z^{(j)}\bigr),
\end{eqnarray*}
where $\operatorname{Hamm}(z^{(i)},z^{(j)})$ is the Hamming distance between sequences
$z^{(i)}$ and $z^{(j)}$ after finding the optimal permutation of the
labels in
test sequence $z^{(i)}$ to those in reference sequence $z^{(j)}$. At a high
level, this method for choosing state sequence samples aims to
produce segmentations of the data that are \textit{typical} samples
from the posterior. %Asymptotically, this approach will minimize a
%posterior expected risk.
\citet{Jasra05} provide an overview of
some related techniques to address the label-switching issue.
Although we could have chosen any label-invariant loss function to
minimize, we chose the Hamming distance metric because it is closely
related to the official NIST \textit{diarization error rate} (DER)
that is calculated during the evaluations. The final metric by which
the speaker diarization algorithms are judged is the \textit{overall}
DER, a weighted average over the set of meetings based on the length
of each meeting.

%f12 ###
\begin{figure*}[t]

\includegraphics{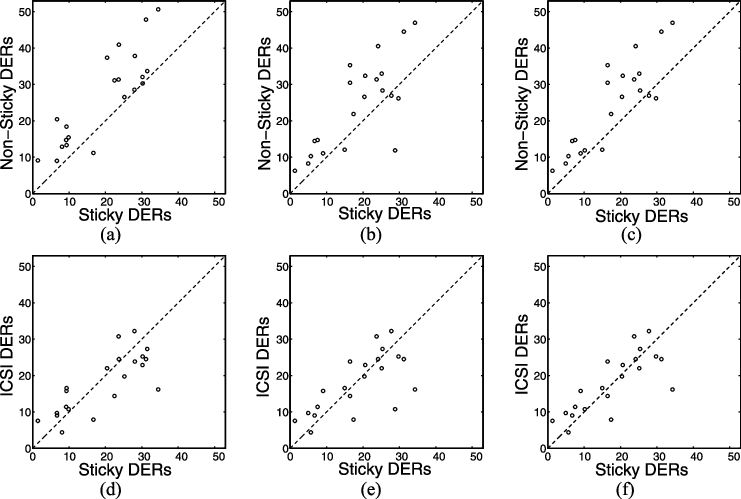}

\caption{\textup{(a)--(c)} For each of the {21} meetings, comparison of diarizations
using sticky vs. original HDP-HMM with DP emissions. In \textup{(a)} we plot the DERs
corresponding to the Viterbi state sequence using the parameters
inferred at Gibbs iteration {10,000} that maximize the likelihood, and in
\textup{(b)} the DERs using the state sequences that minimize the expected Hamming
distance. Plot \textup{(c)} is the same as (\textup{b)}, except for running the {10} chains
for meeting {16} out to {50,000} iterations. \textup{(d)--(f)} Comparison of the sticky
HDP-HMM with DP emissions to the ICSI errors under the same
conditions.} \label{fig:DER}
\end{figure*}

%f13 ###
\begin{figure*}[t]

\includegraphics{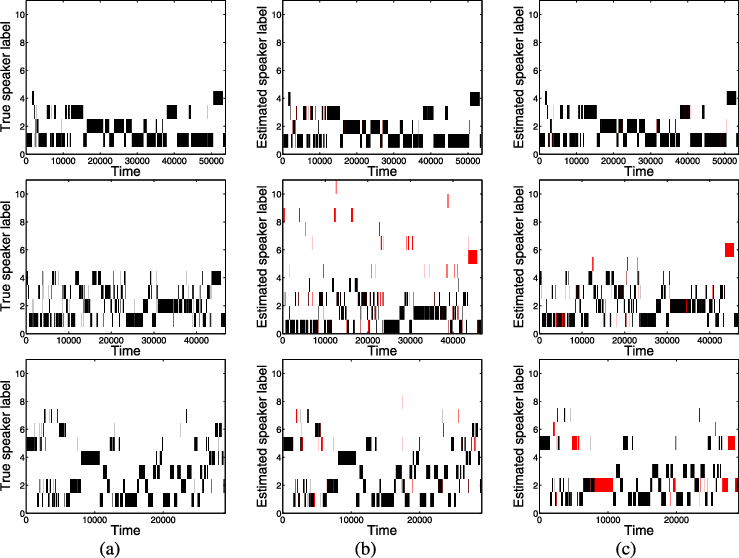}
\vspace*{-5pt}
\caption{Qualitative results for meetings AMI\_\textit{20041210-1052} (meeting
{1}, top), CMU\_\textit{20050228-1615} (meeting {3}, middle) and NIST\_\textit{20051102-1323} meeting (meeting {16}, bottom).
\textup{(a)} True state sequence with the post-processed regions of
overlapping- and nonspeech time steps removed. \textup{(b)} and \textup{(c)}
Plotted only over the time-steps as in \textup{(a)},
the state sequences inferred by the sticky HDP-HMM with DP emissions at Gibbs
iteration {10,000} chosen using the most likely and minimum
expected Hamming distance metrics, respectively. Incorrect labels are
shown in red. For meeting {1}, the maximum likelihood and minimum
expected Hamming distance diarizations are similar, whereas in meeting
{3} we clearly see the sensitivity of the maximum likelihood metric to
overfitting. The minimum expected Hamming distance diarization for
meeting {16} has more errors than that of the maximum likelihood due to
poor mixing rates and many samples failing to identify a~speaker.}
\label{fig:diarizations}\vspace*{-5pt}
\end{figure*}

In Figure~\ref{fig:DER}(a) we report the DER of the chain with the
largest likelihood given the parameters estimated at the
10,000{th} Gibbs iteration for each of the 21 meetings, comparing
the sticky and original HDP-HMM with DP emissions. We see that the
sticky model's temporal smoothing provides substantial performance
gains. Although not depicted in this paper, the likelihoods based on
the parameter estimates under the original HDP-HMM are almost always
higher than those under the sticky model. This phenomenon is due to
the fact that without the sticky parameter, the HDP-HMM
over-segments the data and thus produces parameter estimates more
finely tuned to the data, resulting in higher likelihoods. Since the
original HDP-HMM is contained within the class of sticky models
(i.e., when $\kappa=0$), there is some probability that state
sequences similar to those under the original model will eventually
arise using the sticky model. Thus, since the parameters associated
with these fast-switching sequences result in higher likelihood of the
data, the likelihood metric is not
very robust---one would expect the performance under the sticky
model to degrade given enough Gibbs chains and/or iterations. In
Figure~\ref{fig:DER}(b) we instead report the DER of the chain whose
state sequence estimate at Gibbs iteration 10,000 (this defines the
test set~$\mathcal{T}$)
minimizes the
expected Hamming distance to the sequences estimated every 100 Gibbs
iteration, discarding the first 5000 iterations (this defines the
reference set $\mathcal{R}$). Due to the slow
mixing rate of the chains in this application, we additionally
discard samples whose normalized log-likelihood is below 0.1 units
of the maximum at Gibbs iteration 10,000. From this figure, we see
that the sticky model still significantly outperforms the original
HDP-HMM, implying that most state sequences produced by the original
model are worse, not just the one corresponding to the most likely
sample. Example maximum likelihood and minimum expected Hamming
distance diarizations are displayed in Figure~\ref{fig:diarizations}.
One noticeable exception to this trend is the
NIST\_20051102-1323 meeting (meeting 16). For the sticky model, the
state sequence using the maximum likelihood metric had very low DER
[see Figure~\ref{fig:diarizations}(b)]; however, there were many
chains that merged speakers and produced segmentations similar to
the one in Figure~\ref{fig:diarizations}(c), resulting in such a
sequence minimizing the expected Hamming distance. See
Section~\ref{sec:discussion} for a discussion on the issue of merged
speakers. Running meeting 16 for 50,000 Gibbs iterations improved
the performance, as depicted by the revised results in
Figure~\ref{fig:DER}(c). We summarize our overall performance in
Table~\ref{table:DERs}, and note that (when using the 50,000 Gibbs
iterations for meeting 16 and 10,000 Gibbs iterations for all other
meetings\footnote{On such a large data set, running 10 chains for
50,000 iterations for each of the 21 meetings would have represented a
significant computational burden and, thus, we only ran the chains to
50,000 iterations for meeting 16, which clearly had not mixed after
10,000 iterations (based on an examination of trace plots of
log-likelihoods; see Figure~\ref{fig:meeting16traces}). In meeting 16
the differences between two of the speakers are especially subtle, and
our sampler has difficulty in reliably finding parameters that separate
these speakers.}) we obtain an overall DER of 17.84\% using
the sticky HDP-HMM versus the 23.91\% of the original HDP-HMM model.
Alternatively, when constrained to single Gaussian emissions the sticky
HDP-HMM and original HDP-HMM have overall DERs of 34.97\% and 36.89\%,
respectively, which clearly demonstrates the importance of considering
DP emissions. When considering the DP mixture-of-Gaussians model
(ignoring the time indices associated with the observations), the
overall DER is 72.67\%. If one uses the ground truth labels to map
multiple inferred DP mixture components to a single speaker label, the
overall DER drops to 54.19\%. %This DP mixture of Gaussian many-to-one
%mapping result aims to explore the extent to which speakers, each
%modeled as a mixture of Gaussians, can be identified without relying
%on the temporal dynamics.
The poor performance of the DP mixture-of-Gaussians model, %even when
%assuming ground truth labels are available to employ such a
%many-to-one mapping,
even when assuming that ground truth labels are available, which would
not be the case in practice, illustrates the importance of the temporal
dynamics captured by the HMM.

%t1 ###
\begin{table}
\caption{Overall DERs for the sticky and original
HDP-HMM with DP emissions using the minimum expected Hamming
distance and maximum likelihood metrics for choosing state
%\break
sequences
at Gibbs iteration 10,000}\label{table:DERs}%
\vspace*{-5pt}
\begin{tabular*}{\tablewidth}{@{\extracolsep{4in minus 4in}}lcccc@{}}
\hline
\textbf{Overall DERs (\%)} & \textbf{Min Hamming} & \textbf{Max likelihood} & \textbf{2-Best} & \textbf{5-Best}\\
\hline
Sticky HDP-HMM & 19.01 (17.84) & 19.37 & 16.97 & 14.61 \\
Nonsticky HDP-HMM & 23.91 \phantom{(17.84)} & 25.91 & 23.67 & 21.06 \\
\hline
\end{tabular*}
\tabnotetext[]{}{\textit{Notes}: For the maximum likelihood criterion, we
show the best overall DER if we consider the top two or top five
most likely candidates. The number in the parentheses is the
performance when running meeting 16 for 50,000 Gibbs iterations. The
overall ICSI DER is 18.37\%, while the best achievable DER with the
chosen acoustic preprocessing is 10.57\%.}\vspace*{-2pt}
\end{table}

As a further comparison, the algorithm that was by far the best
performer at the 2007 NIST competition---the algorithm developed by
a team at the International Computer Science Institute
(ICSI)~[\citet{Wooters07}]---has an overall DER of 18.37\%. The ICSI
team's algorithm uses agglomerative clustering, and requires
significant tuning of parameters on representative training data. In
contrast, our hyperparameters are automatically set
meeting-by-meeting, as outlined at the beginning of this section. An
additional benefit of the sticky HDP-HMM over the ICSI approach is
the fact that there is inherent posterior uncertainty in this task,
and by taking a Bayesian approach, we are able to provide several
interpretations. Indeed, when considering the best per-meeting DER
for the five most likely samples, our overall DER drops to 14.61\%
(see Table~\ref{table:DERs}). Although not helpful in the NIST
evaluations, which require a single segmentation, providing multiple
segmentations could be useful in
practice.

%f14 ###
\begin{figure*}[t]

\includegraphics{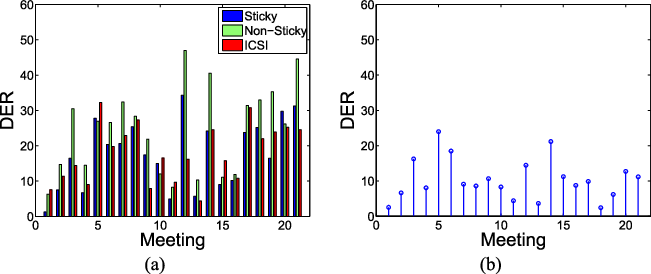}
\vspace*{-5pt}
\caption{\textup{(a)} Chart comparing the DERs of the sticky and original HDP-HMM
with DP emissions to those of ICSI for each of the {21} meetings.
Here, we chose the state sequence at the {10,000}{th} Gibbs iteration
that minimizes the expected Hamming distance. For meeting 16 using
the sticky HDP-HMM with DP emissions, we chose between state sequences at
Gibbs iteration {50,000}. \textup{(b)} DERs associated with using ground truth
speaker labels for the post-processed data. Here, we assign undetected
nonspeech a label different than the preprocessed nonspeech.}
\label{fig:meeting_by_meeting_DERs}\vspace*{-5pt}
\end{figure*}

To ensure a fair comparison, we use the same speech/nonspeech
preprocessing and acoustic features as ICSI, so that the differences in
our performance
are due to changes in the identified speakers. As depicted in
Figure~\ref{fig:meeting_by_meeting_DERs},
both our performance and that of ICSI depend significantly on the
quality of this preprocessing step. For the periods of nonspeech that
are incorrectly identified as speech during preprocessing, we are
forced to produce errors on these sections since they will be assigned
an HMM label (and thus a~speaker label) that is separate from the label
assigned to the preprocessed sections labeled as nonspeech. Another
source of errors are periods of overlapping speech, which impede our
ability to clearly identify a single speaker. In
Figure~\ref{fig:meeting_by_meeting_DERs}(a) we compare the
meeting-by-meeting DERs of the sticky HDP-HMM, the original HDP-HMM,
and the ICSI algorithm. If we use the ground truth speaker labels for
the post-processed data (assigning undetected nonspeech a label
different than the preprocessed nonspeech), the resulting overall DER
is 10.57\% with meeting-by-meeting DERs displayed in Figure~\ref
{fig:meeting_by_meeting_DERs}(b). This number provides a lower bound on
the achievable performance using the speech/nonspeech preprocessing,
our block-averaging of features, and our assumptions of minimum
duration. Beyond these forced errors, it is clear from Figure~\ref
{fig:meeting_by_meeting_DERs}(a) that the sticky
HDP-HMM with DP emissions provides performance comparable to that of
the ICSI algorithm, while the original HDP-HMM with DP emissions
performs significantly worse. Overall, the results presented in
this section demonstrate that the sticky HDP-HMM with DP emissions
provides an elegant and empirically effective speaker diarization
method.

%%
% \centering
% \begin{tabular}{cccc}
% \hspace{-0.2in}\includegraphics[height =
%1.25in]{JournalFigs/truestateseq1} & \hspace{-0.3in}
% \hspace{-0.2in}(a) & \hspace{-0.3in}(b) & \hspace{-0.3in}(c) &
% \hspace{-0.3in}(d)\vspace{-0.1in}
% \end{tabular}
% \caption{True state sequences for meetings (a) AMI\_20041210-1052 and
%(c) VT\_20050304-1300, with the corresponding most likely state
%estimates shown in (b) and (d), respectively, with incorrect labels
%shown in red.} \label{fig:diarizations}
%%
% \centering
% \begin{tabular}{cc}
%
%
%& \hspace{0.25in}\includegraphics[height =
%1.25in]{JournalFigs/scatternumspeak10}\vspace{-0.05in}\\
% (a) & \hspace{0.25in} (b)
% \vspace{-0.1in}
% \end{tabular}
% \caption{For the 21 meeting database: (a) plot of sticky vs. original
%HDP-HMM most
%likely sequence DER; and (b) plot of true vs. estimated number of
%speakers for samples drawn from 10 random initializations of each
%meeting (larger circles have higher likelihood).} \label{fig:DER}
%%

%s9 ###
\section{Discussion}
\label{sec:discussion}
We have developed a Bayesian nonparametric approach to the problem
of speaker diarization, building on the HDP-HMM presented
in \citet{Teh06}. Although the original HDP-HMM does not yield
competitive speaker diarization performance due to its inadequate
modeling of the temporal persistence of states, the sticky HDP-HMM
that we have presented here resolves this problem and yields a
state-of-the-art solution to the speaker diarization problem.

We have also shown that this sticky HDP-HMM allows a fully Bayesian
nonparametric treatment of multimodal emissions, disambiguated by
its bias toward self-transitions. Accommodating multimodal
emissions is essential for the speaker diarization problem and is
likely to be an important ingredient in other applications of the
HDP-HMM to problems in speech technology.

\begin{figure*}[t]

\includegraphics{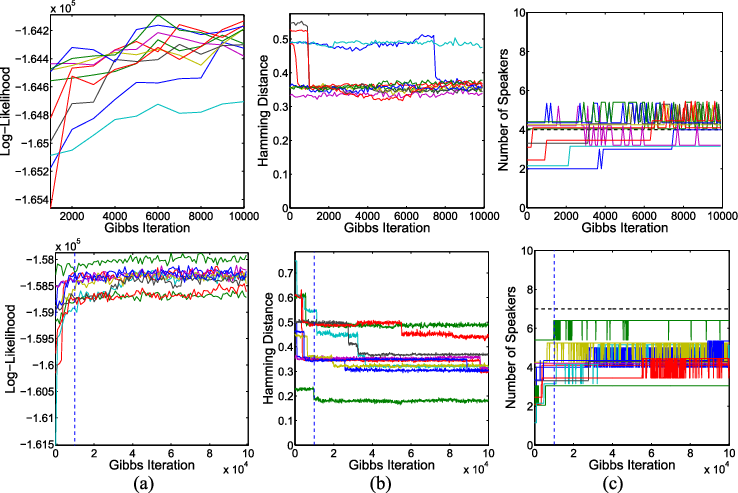}
\vspace*{-5pt}
\caption{Trace plots of \textup{(a)} log-likelihood, \textup{(b)} Hamming distance
error and \textup{(c)} number of speakers
for {10} chains for two meetings: CMU\_\textit{20050912-0900} / meeting {5} (top)
and NIST\_\textit{20051102-1323} / meeting \textit{16} (bottom). For meeting {5}, which has
behavior representative of the majority of the meetings, we show traces
over the {10,000} Gibbs iterations used for the results in Section~\protect\ref
{sec:SpeakerDiarization}. For meeting {16}, we ran the chains out to
{100,000} Gibbs iterations to demonstrate the especially slow mixing rate
for this meeting. The dashed blue vertical lines indicate {10,000}
iterations.} \label{fig:meeting16traces}%
\vspace*{-6pt}
\end{figure*}

We also presented efficient sampling techniques with mixing rates
that improve on the state of the art by harnessing the Markovian
structure of the HDP-HMM. Specifically, we proposed employing a
truncated approximation to the HDP and block-sampling the state
sequence using a variant of the forward--backward
algorithm. Although the blocked samplers %of Algorithms~
yield substantially improved mixing rates over the sequential,
direct assignment samplers, %of Algorithms~\ref{alg:direct} and
there are still some pitfalls to these sampling methods. One issue
is that for each new considered state, the parameter sampled from
the prior distribution must better explain the data than the
parameters associated with other states that have already been
informed by the data. In high-dimensional applications, and in cases
where state-specific emission distributions are not clearly
distinguishable, this method for adding new states poses a
significant challenge.
%The data in the speaker diarization task is both high-dimensional and
%often has only marginally distinguishable speakers, leading to
%extremely slow mixing rates, as indicated by the trace plots in Fig.~
%speaker counts for 10,000 Gibbs sampling iterations of meeting 5 and
%100,000 iterations of meeting 16. We see rather infrequent changes in
%the number of states, or speakers, which is indicative of the mixing
%problems that arise from the challenges in discovering new states in
%this high-dimensional dataset. Many of our errors in this application
%can be attributed to merged speakers, as depicted in Fig.~
%cost of running hundreds of thousands of Gibbs iterations proves an
%insurmountable barrier. A direction for future work is to develop
%split-merge algorithms for the HDP and HDP-HMM similar to those
%developed in~\citet{Jain04} for the DP mixture model.
Indeed, both issues arise in the speaker diarization task and we did
have difficulties with mixing. Further evidence of this is presented in
the trace plots in Figure~\ref{fig:meeting16traces}, where we plot
log-likelihoods, Hamming distances and speaker counts for 10,000 Gibbs
sampling iterations of meeting 5 and
100,000 iterations of meeting 16. As discussed previously, meeting 16
is the most problematic meeting in our data set, and these plots
provide clear evidence that our sampler is not mixing on this meeting.
But even on meeting 5, which is more representative of the full set of
meetings and which is segmented effectively by our procedure, we see a
relatively slow evolution of the sampler, particularly as measured by
the number of speakers. Our use of the minimum expected Hamming
distance procedure to select samples mitigates this difficulty, but
further work on sampling procedures for the sticky HDP-HMM is needed.
One possibility is to consider split-merge algorithms similar to those
developed in~\citet{Jain04} for the DP mixture model.

%f15 ###

A limitation of the HMM in general is that the observations are
assumed conditionally i.i.d. given the state sequence. This
assumption is often insufficient in capturing the complex temporal
dependencies exhibited in real-world data. Another area of future
work is to consider Bayesian nonparametric versions of models better
suited to such applications, like the switching linear dynamical
system (SLDS) and switching VAR process. A first attempt at
developing such models is presented in~\citet{FoxNIPS08}. An
inspiration for the sticky HDP-HMM actually came from considering
the original HDP-HMM as a prior for an SLDS. In such scenarios where
one does not have direct observations of the underlying state
sequence, the issues arising from not properly capturing state
persistence are exacerbated. The sticky HDP-HMM presented in this
paper provides a robust building block for developing more
complex Bayesian nonparametric dynamical models.

\section*{Acknowledgments}
We thank O. Vinyals, G. Friedland and N. Morgan for helpful
discussions about the NIST data set.

\begin{supplement}[id=suppA]
\sname{Supplement}
\stitle{Notational conventions, Chinese restaurant
franchises and derivations of Gibbs samplers}
\slink[doi]{10.1214/10-AOAS395SUPP}
\slink[url]{http://lib.stat.cmu.edu/aoas/395/supplement.pdf}
\sdatatype{.pdf}
\sdescription{We present detailed derivations of the
conditional distributions used for both the direct assignment and blocked
Gibbs samplers, as well as the associated pseudo-code.  The description of
these derivations relies on the Chinese restaurant analogies associated with
the HDP and sticky HDP-HMM, which are expounded\vadjust{\goodbreak} upon in this supplementary
material.  We also provide a list of notational conventions used throughout
the paper.}
\end{supplement}

%suskaldyti doi

%suskaldyti doi

\printaddresses


\begin{thebibliography}{42}
% \providecommand{\doi}[1]{doi: #1}\else
% \providecommand{\doi}{doi: \begingroup\urlstyle{rm}\Url}\fi

%%b1 ###
% \textsc{Antoniak}, C. E. (1974).
%Mixtures of {D}irichlet processes with applications to {B}ayesian
%nonparametric problems.


%b2 ###
\bibitem[\protect\citeauthoryear{Barras et~al.}{2004}]{Barras04}
 \textsc{Barras}, C.,  \textsc{Zhu}, X.,  \textsc{Meignier}, S. and  \textsc{Gauvain}, J.-L.
 (2004). Improving speaker diarization.
In \textit{Proc. Fall 2004 Rich Transcription Workshop (RT-04)},
November 2004.

%b3 ###
\bibitem[\protect\citeauthoryear{Beal and Krishnamurthy}{2006}]{Beal06}
 \textsc{Beal}, M. J. and  \textsc{Krishnamurthy}, P. (2006).
Gene expression time course clustering with countably infinite hidden
{M}arkov models.
In \textit{Proc. Conference on Uncertainty in Artificial Intelligence},
Cambridge, MA.

%b4 ###
\bibitem[\protect\citeauthoryear{Beal, Ghahramani and Rasmussen}{2002}]{Beal02}
 \textsc{Beal}, M. J.,  \textsc{Ghahramani}, Z. and  \textsc{Rasmussen}, C. E. (2002).
The infinite hidden {M}arkov model.
In \textit{Advances in Neural Information Processing Systems}
 \textbf{14}  577--584. MIT Press, Cambridge, MA.

%b5 ###
\bibitem[\protect\citeauthoryear{Blackwell and MacQueen}{1973}]{Blackwell73}
 \textsc{Blackwell}, D. and \textsc{MacQueen}, J. B. (1973).
Ferguson distributions via {P}\'{o}lya urn schemes.
\textit{Ann. Statist.} \textbf{1}  353--355.
\MR{0362614}


%%b6 ###
% \textsc{Casella}, G. and  \textsc{Robert}, C. (1996).
%Rao-{B}lackwellisation of sampling schemes.

%b7 ###
\bibitem[\protect\citeauthoryear{Chen and Gopalakrishnam}{1998}]{Chen98}
 \textsc{Chen}, S. S. and  \textsc{Gopalakrishnam}, P. S. (1998).
Speaker, environment and channel change detection and clustering via
the {B}ayesian information criterion.
In \textit{Proc. {DARPA} Broadcast News Transcription and Understanding
Workshop} 127--132. Morgan Kaufmann, San Francisco, CA.

%%b8 ###
% \textsc{Escobar}, M. D. and  \textsc{West}, M. (1995).
%Bayesian density estimation and inference using mixtures.
% 577--588.

%b9 ###
\bibitem[\protect\citeauthoryear{Ferguson}{1973}]{Ferguson73}
 \textsc{Ferguson}, T. S. (1973).
A {B}ayesian analysis of some nonparametric problems.
\textit{Ann. Statist.} \textbf{1}  209--230.
\MR{0350949}


%b10 ###
\bibitem[\protect\citeauthoryear{Fox et~al.}{2008}]{FoxICML08}
 \textsc{Fox}, E. B.,  \textsc{Sudderth}, E. B.,
 \textsc{Jordan}, M. I. and  \textsc{Willsky}, A. S. (2008).
An {HDP-HMM} for systems with state persistence.
In \textit{Proc. International Conference on Machine Learning}, Helsinki, Finland, July
2008.


%b11 ###
\bibitem[\protect\citeauthoryear{Fox et~al.}{2009}]{FoxNIPS08}
 \textsc{Fox}, E. B.,  \textsc{Sudderth}, E. B.,
 \textsc{Jordan}, M. I. and  \textsc{Willsky}, A. S. (2009).
Nonparametric {B}ayesian learning of switching dynamical systems.
In \textit{Advances in Neural Information Processing Systems}
 \textbf{21}  457--464.

%b12 ###
\bibitem[\protect\citeauthoryear{Fox et~al.}{2010}]{SuppA}
 \textsc{Fox}, E. B.,  \textsc{Sudderth}, E. B.,  \textsc{Jordan}, M. I. and  \textsc{Willsky}, A. S. (2010).
Supplement to ``A sticky HDP-HMM with application to speaker diarization.''
DOI:
\href{http://dx.doi.org/10.1214/10-AOAS395SUPP}{10.1214/10-AOAS395SUPP}.

%b13 ###
\bibitem[\protect\citeauthoryear{Gales and Young}{2007}]{GalesYoung}
 \textsc{Gales}, M. and  \textsc{Young}, S. (2007).
{The Application of hidden Markov models in speech recognition}.
\textit{Foundations and Trends in Signal Processing} \textbf{1} 195--304.



%b14 ###
\bibitem[\protect\citeauthoryear{Gauvain, Lamel and Adda}{1998}]{Gauvain98}
 \textsc{Gauvain}, J.-L.,  \textsc{Lamel}, L. and  \textsc{Adda}, G. (1998).
Partitioning and transcription of broadcast news data.
In \textit{Proc. International Conference on Spoken Language
Processing}, Sydney, Australia 1335--1338.

%%b15 ###
% \textsc{Gelman}, A.,  \textsc{Carlin}, J. B.,  \textsc{Stern}, H. S. and  \textsc{Rubin}, D. B. (2004).
%Chapman \& Hall.

%b16 ###
\bibitem[\protect\citeauthoryear{Hoffman, Cook and Blei}{2008}]{Hoffman08}
 \textsc{Hoffman}, M.,  \textsc{Cook}, P. and  \textsc{Blei}, D. (2008).
Data-driven recomposition using the hierarchical {D}irichlet process
hidden {M}arkov model.
In \textit{Proc. International Computer Music Conference}, Belfast, UK.

%b17 ###
\bibitem[\protect\citeauthoryear{Ishwaran and Zarepour}{2000a}]{Ishwaran00}
 \textsc{Ishwaran}, H. and  \textsc{Zarepour}, M. (2000a).
Markov chain {M}onte {C}arlo in approximate {D}irichlet and beta
two--parameter process hierarchical models.
\textit{Biometrika} \textbf{87}  371--390.
\MR{1782485}

%b18 ###
\bibitem[\protect\citeauthoryear{Ishwaran and Zarepour}{2002b}]{Ishwaran02}
 \textsc{Ishwaran}, H. and  \textsc{Zarepour}, M. (2002b).
Dirichlet prior sieves in finite normal mixtures.
\textit{Statist. Sinica} \textbf{12} 941--963.
\MR{1929973}

%b19 ###
\bibitem[\protect\citeauthoryear{Ishwaran and
Zarepour}{2002c}]{Ishwaran02-2}
 \textsc{Ishwaran}, H. and  \textsc{Zarepour}, M. (2002c).
Exact and approximate sum---representations for the {D}irichlet
process.
\textit{Canad. J. Statist.} \textbf{30} 269--283.
\MR{1926065}

%b20 ###
\bibitem[\protect\citeauthoryear{Jain and Neal}{2004}]{Jain04}
 \textsc{Jain}, S. and  \textsc{Neal}, R. M. (2004).
A split-merge Markov chain Monte Carlo procedure for the dirichlet
process mixture model.
\textit{J. Comput. Graph. Statist.}
\textbf{13} 158--182.
\MR{2044876}

%b21 ###
\bibitem[\protect\citeauthoryear{Jasra, Holmes and Stephens}{2005}]{Jasra05}
 \textsc{Jasra}, A.,  \textsc{Holmes}, C. C. and  \textsc{Stephens}, D. A. (2005).
Markov chain {M}onte {C}arlo methods and the label switching problem
in {B}ayesian mixture modeling.
\textit{Statist. Sci.} \textbf{20}  50--67.
\MR{2182987}

%b22 ###
\bibitem[\protect\citeauthoryear{Johnson}{2007}]{Johnson07}
 \textsc{Johnson}, M. (2007).
Why doesn't {EM} find good {HMM} {POS}-taggers.
In \textit{Proc. Joint Conference on Empirical Methods in Natural
Language Processing and Computational Natural Language Learning},
Prague, Czech Republic.\

%b23 ###
\bibitem[\protect\citeauthoryear{Kivinen, Sudderth and
Jordan}{2007}]{Kivinen07}
 \textsc{Kivinen}, J. J.,  \textsc{Sudderth}, E. B. and  \textsc{Jordan}, M. I. (2007).
Learning multiscale representations of natural scenes using
{D}irichlet processes.
In \textit{Proc. International Conference on Computer Vision},
Rio de Janeiro, Brazil 1--8.

%b24 ###
\bibitem[\protect\citeauthoryear{Kurihara, Welling and
Teh}{2007}]{Kurihara07}
 \textsc{Kurihara}, K.,  \textsc{Welling}, M. and  \textsc{Teh}, Y. W. (2007).
Collapsed variational {D}irichlet process mixture models.
In \textit{Proc. International Joint Conferences on Artificial
Intelligence}, Hyderabad, India.

%b25 ###
\bibitem[\protect\citeauthoryear{Meignier et~al.}{2000}]{Meignier00}
 \textsc{Meignier}, S.,  \textsc{Bonastre}, J.-F.,  \textsc{Fredouille}, C. and  \textsc{Merlin}, T. (2000).
Evolutive {HMM} for multi-speaker tracking system.
In \textit{Proc. IEEE International Conference on Acoustics, Speech and
Signal Processing (ICASSP)}, Istanbul, Turkey, June 2000.

%b26 ###
\bibitem[\protect\citeauthoryear{Meignier, Bonastre and
Igounet}{2001}]{Meignier01}
 \textsc{Meignier}, S.,  \textsc{Bonastre}, J.-F. and  \textsc{Igounet},
 S. (2001).
{E-HMM} approach for learning and adapting sound models for speaker
indexing.
In \textit{Proc. Odyssey Speaker Language Recognition Workshop}, June
2001.

%b27 ###
\bibitem[\protect\citeauthoryear{Munkres}{1957}]{Munkres57}
 \textsc{Munkres}, J. (1957).
Algorithms for the assignment and transportation problems.
\textit{J. Soc. Industr. Appl. Math.}
\textbf{5}  32--38.
\MR{0093429}

%b28 ###
\bibitem[\protect\citeauthoryear{NIST}{2007}]{NIST}
NIST.
Rich transcriptions database. Available at
\href{http://www.nist.gov/speech/tests/rt/}{http://www.nist.gov/speech/tests/rt/}, 2007.

%b29 ###
\bibitem[\protect\citeauthoryear{Papaspiliopoulos and
Roberts}{2008}]{Papaspiliopoulos08}
 \textsc{Papaspiliopoulos}, O. and  \textsc{Roberts}, G. O. (2008).
Retrospective {M}arkov chain {M}onte {C}arlo methods for {D}irichlet
process hierarchical models.
\textit{Biometrika} \textbf{95} 169--186.
\MR{2409721}

%b30 ###
\bibitem[\protect\citeauthoryear{Rabiner}{1989}]{Rabiner89}
 \textsc{Rabiner}, L. R. (1989).
A tutorial on hidden {M}arkov models and selected applications in
speech recognition.
\textit{Proc. IEEE} \textbf{77}  257--286.


%b31 ###
\bibitem[\protect\citeauthoryear{Reynolds and
Torres-Carrasquillo}{2004}]{Reynolds04}
 \textsc{Reynolds}, D. A. and \textsc{Torres-Carrasquillo}, P. A. (2004).
The {MIT} {L}incoln {L}aboratory {RT-04F} diarization systems:
{A}pplications to broadcast news and telephone conversations.
In \textit{Proc. Fall 2004 Rich Transcription Workshop (RT-04)},
November 2004.

%b32 ###
\bibitem[\protect\citeauthoryear{Robert}{2007}]{Robert}
 \textsc{Robert}, C. P. (2007).
\textit{The {B}ayesian Choice}.
Springer, New York.

%b33 ###
\bibitem[\protect\citeauthoryear{Rodriguez, Dunson and
Gelfand}{2008}]{Rodriguez08}
 \textsc{Rodriguez}, A.,  \textsc{Dunson}, D. B. and  \textsc{Gelfand}, A. E. (2008).
The nested {D}irichlet process.
\textit{J.~Amer. Statist. Assoc.} \textbf{103}
 1131--1154.

%b34 ###
\bibitem[\protect\citeauthoryear{Scott}{2002}]{Scott02}
 \textsc{Scott}, S. L. (2002).
Bayesian methods for hidden {M}arkov models: Recursive computing in
the 21st century.
\textit{J. Amer. Statist. Assoc.} \textbf{97}
 337--351.
\MR{1963393}

%b35 ###
\bibitem[\protect\citeauthoryear{Sethuraman}{1994}]{Sethuraman94}
 \textsc{Sethuraman}, J. (1994).
A constructive definition of {D}irichlet priors.
\textit{Statist. Sinica} \textbf{4} 639--650.
\MR{1309433}

%b36 ###
\bibitem[\protect\citeauthoryear{Siegler et~al.}{1997}]{Siegler97}
 \textsc{Siegler}, M.,  \textsc{Jain}, U.,  \textsc{Raj}, B. and  \textsc{Stern}, R. M. (1997).
Automatic segmentation, classification and clustering of broadcast
news audio.
In \textit{Proc. {DARPA} Speech Recognition Workshop} 97--99.
Morgan Kaufmann, San Francisco, CA.


%b37 ###
\bibitem[\protect\citeauthoryear{Teh et~al.}{2006}]{Teh06}
 \textsc{Teh}, Y. W.,  \textsc{Jordan}, M. I.,  \textsc{Beal}, M. J. and  \textsc{Blei}, D. M. (2006).
Hierarchical {D}irichlet processes.
\textit{J. Amer. Statist. Assoc.} \textbf{101}
1566--1581.
\MR{2279480}

%b38 ###
\bibitem[\protect\citeauthoryear{Tranter and Reynolds}{2006}]{Tranter06}
 \textsc{Tranter}, S. E. and  \textsc{Reynolds}, D. A. (2006).
An overview of automatic speaker diarization systems.
\textit{IEEE Trans. Audio, Speech  Language Process.}
\textbf{14} 1557--1565.

%b39 ###
\bibitem[\protect\citeauthoryear{Van~Gael et~al.}{2008}]{VanGael08}
 \textsc{Van Gael}, J.,  \textsc{Saatci}, Y.,  \textsc{Teh}, Y. W. and  \textsc{Ghahramani},
 Z. (2008).
Beam sampling for the infinite hidden {M}arkov model.
In \textit{Proc. International Conference on Machine Learning}, Helsinki, Finland, July
2008.

%b40 ###
\bibitem[\protect\citeauthoryear{Walker}{2007}]{Walker07}
 \textsc{Walker}, S. G. (2007).
Sampling the {D}irichlet mixture model with slices.
\textit{Commun. Statist. Simul. Comput.}
\textbf{36} 45--54.
\MR{2370888}

%b41 ###
\bibitem[\protect\citeauthoryear{Wooters and Huijbregts}{2007}]{Wooters07}
 \textsc{Wooters}, C. and  \textsc{Huijbregts}, M. (2007).
The {ICSI} {RT}07s speaker diarization system.
\textit{Lecture Notes in Computer Science} \textbf{4625} 509--519.

%b42 ###
\bibitem[\protect\citeauthoryear{Wooters et~al.}{2004}]{Wooters04}
 \textsc{Wooters}, C.,  \textsc{Fung}, J.,  \textsc{Peskin}, B. and  \textsc{Anguera}, X. (2004).
Towards robust speaker segmentation: {T}he {ICSI-SRI} {F}all 2004
diarization system.
In \textit{Proc. Fall 2004 Rich Transcription Workshop (RT-04)},
November 2004.\

%b43 ###
\bibitem[\protect\citeauthoryear{Xing and Sohn}{2007}]{Xing07}
 \textsc{Xing}, E. P. and  \textsc{Sohn}, K.-A. (2007).
Hidden {M}arkov {D}irichlet process: Modeling genetic inference in
open ancestral space.
\textit{Bayesian Anal.} \textbf{2}  501--528.
\MR{2342173}

\end{thebibliography}
\end{document}